\documentclass[aps,preprint,showpacs,showkeys,preprintnumbers,amsmath,amssymb]{revtex4-1}

\usepackage{graphicx}
\usepackage{dcolumn}
\usepackage{tabularx}
\usepackage{bm}

\begin{document}

\title{Magnetic, electronic and vibrational properties of metal and fluorinated metal phthalocyanines}

\author{O.\ I.\ Arillo-Flores$^1$}
\author{M.\ M.\ Fadlallah$^{2,3}$}
\author{C. Schuster$^2$}
\author{U. Eckern$^2$}
\author{A.\ H.\ Romero$^{1,4,5}$}\email{aromero@qro.cinvestav.mx}
\affiliation{
$^1$CINVESTAV-Quer\'etaro Libramiento Norponiente No. 2000 Real de Juriquilla
76230 Quer\'etaro, Qro, M\'exico\\
$^2$Institut f\"ur Physik, Universit\"at Augsburg, 86135 Augsburg, Germany\\
$^3$Physics Department, Faculty of Science, Benha University, Benha, Egypt\\
$^4$Max-Planck-Institute f\"ur Mikrostrukturphysik, Weinberg 2, 06120 Halle, Germany\\
$^5$Physics Department, West Virginia University, Morgantown, West Virginia 26506-6315, USA}

\date{\today}

\begin{abstract}
The magnetic and electronic properties of metal phthalocyanines (MPc) and fluorinated metal phthalocyanines (F$_{16}$MPc)
are studied by means of spin density functional theory (SDFT). Several metals (M) such as Ca, all first $d$-row transition
metals and Ag are investigated. 
By considering different open shell transition metals it is possible to tune the
electronic properties of MPc, in particular the electronic molecular gap and total magnetic moment.
Besides assigning the structural and electronic properties of MPc and F$_{16}$MPc, 
the vibrational modes analysis of the ScPc\textendash ZnPc series have been studied and correlated
to experimental measurements when available.

\end{abstract}


\maketitle

\section{Introduction}
Magnetism or magnetic moments are rarely found in organic compounds, but metal phthalocyanine (MPc) molecules are notable exceptions
\cite{je}. The MPc family represents metal-organic semiconductors that display high thermal and
chemical stability \cite{NM}. These molecules have recently attracted large interest 
in applications such as in solar cells, optoelectronic devices, light-emitting diodes,
thin-film transistors, and gas sensors \cite{Va,Pe,Ka,Ba,de}.  The magnetic properties of CoPc are used in spintronic devices \cite{Re3}.
Furthermore, the MPc are simple magnetic molecules which can serve as prototypical systems to study the magnetic properties of the 3$d$ transition metals embedded in an organic surrounding.

Fluorinated metal-phthalocyanines (F$_{16}$MPc) are also widely used in applications. In contrast to
the MPc which usually are p-type like semiconductors, F$_{16}$MPc are used as n-type
 semiconductors in photovoltaic devices and in light-emitting diodes.
Organic field-effect transistors  have been intensively studied as
 components for low price, large area, and flexible circuit applications.
To build them, an n-type organic semiconductor is needed. Experimental investigations focus on F$_{16}$CuPc \cite{HH,TW}
and F$_{16}$CoPc \cite{Ote,Ote1,Scu,Hip}. They have been also considered as
an alternative to inorganic field-effect transistors (FETs) in some specific circuit applications such as radio frequency
identification cards (RFIDs) \cite{Dru,Wat}, electronic paper \cite{Wis}, sensors \cite{Cro,Cro1}, and switching devices for active
matrix flat panel displays (AMFPDs) \cite{Dod}.

Several experimental techniques as X-ray spectroscopy \cite{Brow,Ki,Sche,Eva,Xio}, neutron diffraction \cite{Brow,Ki,Sche,Eva,Xio},
scanning tunneling microscopy (STM) \cite{Hipps,Lip}, nuclear magnetic resonance \cite{Fil}, and photo-emission spectroscopy 
\cite{Schwieger} have been used to characterize the structural and electronic properties of MPc. However, for further 
developments in the synthesis and production of novel materials and devices based on MPc, more accessible and 
common techniques such as IR and Raman spectroscopy are useful. For example,
Raman scattering has already been employed to identify different metal phthalocyanines and to probe polymorphic changes (packing)
\cite{Tackley}, as well as structural modifications \cite{KAN}.
Therefore, a systematic study of vibrational modes for the series MPc will provide information on the
resolution of spectral features obtained with ultraviolet photo-electron,
x-ray photo-emission and photoabsortion spectroscopy
measurements. Vibrational coupling effects have been found to have an important effect in the description of outer molecular orbitals of
CuPc \cite{Eva}, as well as of valence bands and core levels of PbPc films \cite{Papageorgiou}. The study of vibrational
properties can help to improve the assessment of other theoretical methods used to study metal-organic complexes 
\cite{Rosa}, and to correlate magnetic transitions to changes of the surrounding \cite{Gauyacq}.

From a theoretical point of view, density functional theory (DFT) is a widely used method to determine the 
electronic structure of  various organic molecules, including MPc \cite{Me,KAN,LL}. Bialek {\it et al.}
applied an all-electron full-potential linearized augmented plane-wave  method to study
the electronic structure and magnetic moments of NiPc, CuPc, FePc, and CoPc \cite{Bi1,Bi2,Bi3,Bi4}.
Marom {\it et al.} have tested a variety of exchange correlation functionals for copper phthalocyanine \cite{Marom08},
comparing local functionals such as LDA (local density approximation) \cite{R0}
and GGA (generalized gradient approximation) \cite{R01,R02} with the semiempirical hybrid functional B3LYP \cite{R03},
the non-empirical PBE \cite{R1}, PBE hybrid (PBEh) \cite{R2},  and the screened HSE \cite{R3}.
While all these functionals describe the geometrical structure quite well,
the local functionals underestimate the binding of orbitals at the molecule center.
Additionally,  GGA+U calculations have been performed in Ref.\ \cite{Bhatt} to take into account
the strong correlations that could be relevant around the metal center.

The present article is organized as follows. After a brief description of the computational method in Section II,
we report in Section  III our results for the magnetic and electronic structure of CaPc, of the sequence of $3d$ MPc (M $=$ Sc, Ti, V, Cr,
Mn, Fe, Co, Ni, Cu, Zn), and of $4d$ AgPc. In addition, we compare the 
electronic properties of the fluorinated MPc (F$_{16}$MPc) and MPc.
In Section IV we present our results for the vibrational properties of MPc, and we analyze general trends. 
A brief summary is given in Section V.

\section{Computational methods}

In the present work,
we focus on studying the electronic structure near the Fermi energy, $E_{F}$, and the magnetic structure by means of density functional theory.
Since hybrid functionals with a stronger amount of Fock exchange tend to overestimate magnetic moments and magnetic order, 
we apply the simplest exchange correlation functional, namely the LDA,
as implemented in the SIESTA package \cite{Sol}.  The electronic structure and the vibrational modes 
are calculated in MPc and F$_{16}$MPc by using this implementation.
The calculations have been performed by considering a finite spin polarization to allow for
the formation of local magnetic moments. The PBE method \cite{R1} is also employed in order to compare with the density of states (DOS)  evaluated with LDA.

Wavefunctions in SIESTA are described by a local atomic orbitals basis set; we have used
a double zeta basis set and an energy cutoff of 300 Ry. The pseudo potentials are norm-conserving in fully non-local form.
Structural optimizations were performed using the conjugate gradient method until net force on every atom was smaller than 0.04 eV/\AA.
Non-symmetrical constraints were employed.
At the optimized geometries, force constants were evaluated and employed to calculate vibrational modes and frequencies through the dynamical matrix.
As we are interested in the electronic structure and vibrational properties of single molecules, we restrict our calculations to
the $\Gamma$ point, and a large simulation box of 28 \AA\ $\times$ 27 \AA\ $\times$ 45 \AA\ was used, to avoid periodic images interactions.
Further geometry optimization and frequency calculations of selected systems were also carried out with B3LYP/6-31G**
in the Gaussian 03 suit of packages, in order to identify, by comparison with previous results, the IR and Raman representations.

\section{Magnetic and electronic properties of metal phthalocyanine}

\begin{figure}
\includegraphics[width=0.35\textwidth]{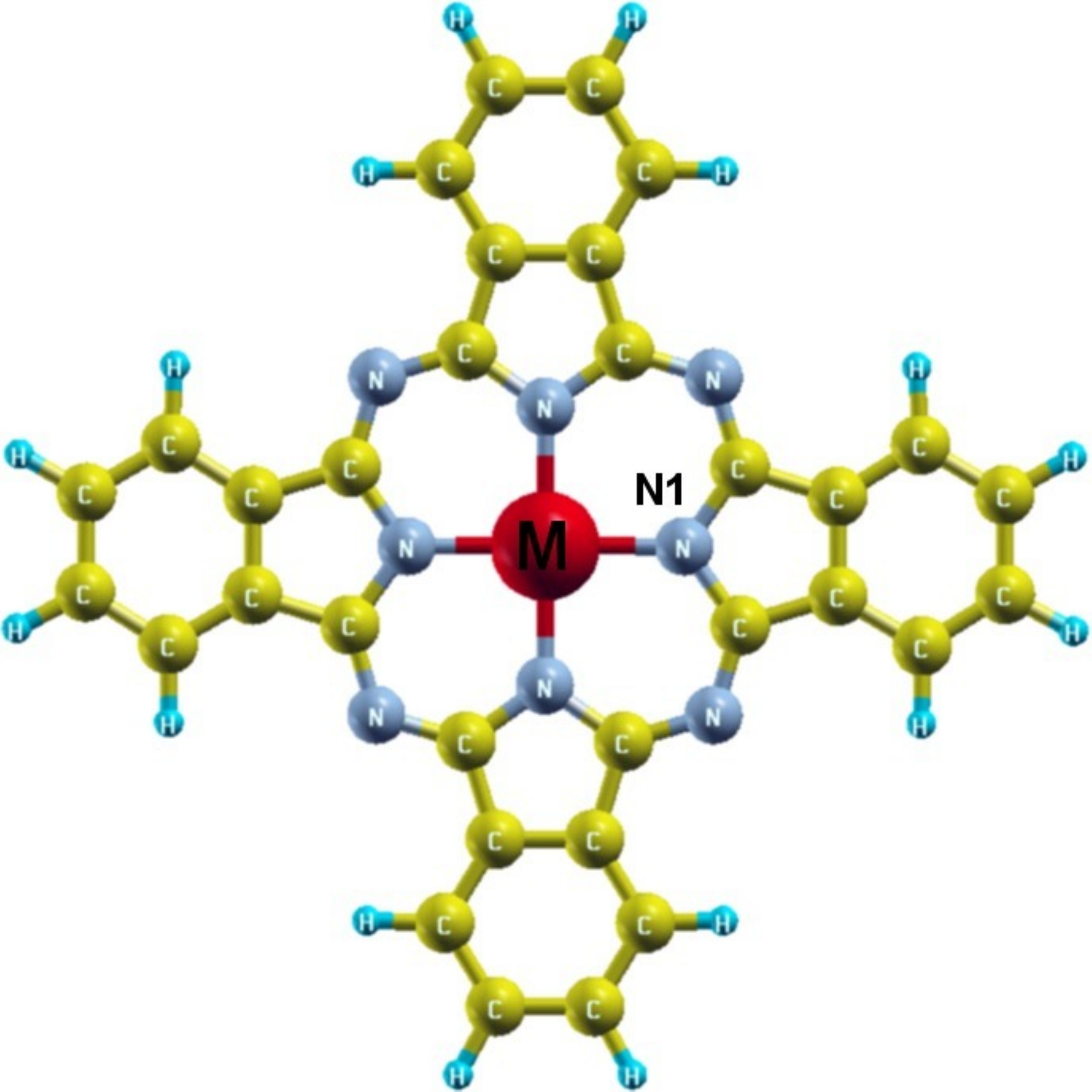}
\caption{(Color online) Structure under investigation: metal phthalocyanine (MPc) molecule. The atoms are labeled according to the chemical constituents, and the metal
atom in the center is denoted by M.}
\label{fig1}
\end{figure}

Metal phthalocyanine, MC$_{32}$N$_{8}$H$_{16}$, is an aromatic molecule with the molecular structure shown in Fig.\ \ref{fig1}.
Hexadecafluorophthalocyaninatometal, F$_{16}$MPc, is constructed by replacing the hydrogen atoms within MPc with fluor atoms.
It is expected that the metal (M) is located at the center of the molecule, displaying a square planar symmetry D$_{4h}$ \cite{Kah}.
The structural, electronic and magnetic properties of MPc are controlled by the particular characteristics of the metal 
within the inner ring. Phenyl groups contribute to the stabilization of these molecular complexes and influence their 
charge distribution, as reported for the formation of ScPc$_{2}$ \cite{Clar,Li}. Nonetheless, they may be subject to 
oxidation as in the case of TiPc \cite{Chen,Li,Toda} and VPc \cite{Barl,Pan}.

To study the magnetic structure of the first row transition metals and also of Ca and of Ag within an isolated MPc, we first focus on the determination of their geometrical structure, and on how our observations relate to our approximations and the different molecular components.

\begin{figure}
\includegraphics[width=0.40\textwidth]{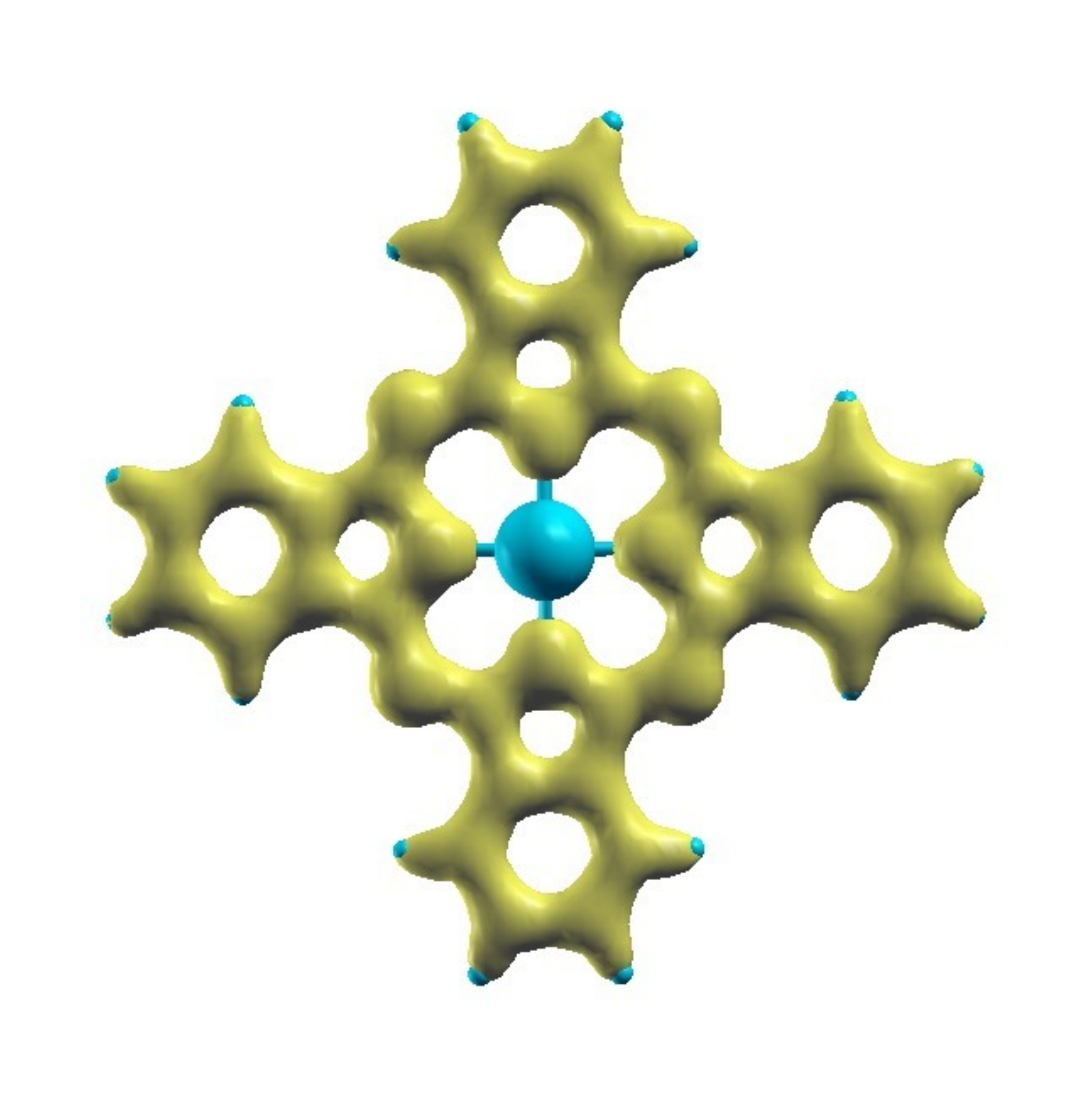}
\caption{ (Color online) Charge density for ScPc. Isosurface value = 0.15}
\label{fig2}
\end{figure}

In general the optimized structures show a planar geometry except for the ScPc and CaPc, where the metal
is slightly out-of-plane while the rest of the molecule undergoes a small bending towards the opposite direction.
This effect becomes larger as the atomic radius of the metal increases, the Sc atom (with radius of 1.62 \AA) is
displaced by 0.24 \AA\ from the molecular plane while Ca (with radius of 1.97 \AA) is shifted by 1.12 \AA.
To confirm the observed trend we calculate the optimized geometry
for YPc obtaining an out-of-plane displacement of 0.72 \AA. 
It can be seen from Fig.\ \ref{fig2} that for ScPc the valence electron density (isosurface value = 0.15) 
appears distant from the metal center, beyond the Van der Waals radius of Sc, i.e., the metal departure from 
the molecular plane is caused by the repulsion between inner electron shells on the metal and on adjacent atoms.
In the distorted geometry the metal is more exposed and the molecular liability increases.

Geometrical structures of minimal energy may be influenced by a Jahn-Teller distortion, which tends
to reduce the symmetry from $D_{4h}$ to $D_{2h}$.
As the point group symmetry was not constrained through our geometry optimizations, strictly speaking
we have obtained, for MPc planar geometries, the $D_{2h}$ symmetry with two M-N distances slightly different
from each other, nevertheless they differ only by less than 1 \%, this variation being in order of magnitude
comparable to the precision of the energy minimization procedure.
A detailed account of the Jahn-Teller distortion effect on the studied MPc series is beyond the scope of the present
work. Interested readers are referred to the work of Marom and Kronik \cite{Marom09b} who emphasize the theoretical and
experimental difficulties to describe precisely the electronic states of MnPc and FePc, the
differences being due to their intricate potential energy surface.

Table \ref{table1} shows the bond lengths between the metal and the nitrogen N1, as indicated in Fig.\ \ref{fig1}.
When available, we compare to experimental data. The theoretical bond lengths differ from experimental lengths by less than 0.05 \AA,
i.e., by 1 to 2 \%.  Decreasing  the atomic covalent radius leads to a decrease in the bond length  from Sc to Cr, as well as from Mn to Cu.
In F$_{16}$MPc, the M--N1 bond length  is typically slightly larger than in  MPc, as shown in Tab.\ \ref{table1}.
The influence of the larger radius of fluorine with respect to hydrogen, and its $p$ electrons show up only when no metallic states are present near the Fermi level.

The last row of Tab.\ \ref{table1} shows the magnetic moments of MPc  (in the case of
F$_{16}$MPc molecules, the calculated magnetic moments are quite close to the related MPc).
Along the sequence from M = Sc to Mn, an open $d$-shell on the metal progressively becomes half-filled and 
the MPc magnetic moments
closely follow the tendency on the metal center (Sc ($s=1/2$, $3d^{1}$), Ti ($s=1$, $3d^{2}$), V ($s=3/2$, $3d^{3}$), 
Cr ($s=2$, $3d^{4}$), and Mn ($s=5/2$, $3d^{5}$)). 
The magnetic moment found for MnPc, $5 \mu_{B}$, is larger than the previously reported value of $3 \mu_{B}$ \cite{Shen},
however, high magnetic moments within this series are also obtained using PBE.
In the second half of the $d$-shell, total spin values lie between 0 and 2. In general, they are smaller than those predicted
for the first half. The values found for this interval agree very well with earlier results \cite{NM,bruder10,zhang11}.
Therefore, the trend in the magnetic moment can be correlated with the electron pairing and the crystal field created 
by the molecule to the metallic atom, which reduces the covalent character.
The atomic orbitals hybridize such that they originate a new mixing, altering the expected orbital ordering from the plain
crystalline field effect, when it goes only over the non-hybridized atomic orbitals.
As a consequence there is an abrupt decrease of magnetic moment from Mn to Fe, and zero magnetic moments are found 
for NiPc before completing the $d$-shell. CuPc ($s=1/2$, $3d^{9}$) and AgPc ($s=1/2$, $4d^{9}$) have a hole in their nearly
full $d$-shell. 
The fact that magnetic moments for F$_{16}$MPc are similar to those for MPc suggests that
molecular magnetic moment are strongly influenced by contributions to the total electronic density coming from the metal center,
with less dependence on the molecular environment; this observation holds for Ag and even Ca, an $s$-block element.

The observed tendency for M--N1 bond lengths results from two main factors. The first one is the metal atomic radius reduction
along the period, which clearly dominates from Ca to V, and the second one is the increase of electronic repulsion for high spin values and degree of orbital occupancy;
these effects overcome the first one for Cr. In the second half of the $d$-row, the decrease in magnetic moment values 
is reflected in the
M--N1 bond lengths variation from Fe to Cu. Deviation from such behavior in the case of ZnPc may be understood by
electron accumulation on the metal when the $d$-shell is full, and the resulting increase of electronic repulsion.

\begin{table}[h]
\begin{center}
\caption{M-N1 bond lengths (\AA) and magnetic moments ({\bf m} ($\mu_{B}$)) for MPc and F$_{16}$MPc} 
\centering
\begin{tabular}{|c|c|c|c|c|c|c|c|c|c|c|c|c|c|c|}
\hline\hline
 M &&Ca& Sc& Ti& V& Cr& Mn&Fe& Co& Ni& Cu& Zn& Ag
\\ [0.5ex]
\hline
&MPc &2.322 &1.998 & 1.974 & 1.968 &1.971 & 2.012 &1.912 &1.883&1.875
&1.937  &1.975&2.042
 \\[-1ex]
&& & &  & & &  &1.927$^a$ &1.912$^b$&1.83$^c$
& 1.935$^d$ &1.980$^e$&
 \\[-1ex]
&& & &  & & &  & &&
&  & 1.954$^f$&
 \\[-1ex]
\raisebox{5.5ex}{Bond}&F$_{16}$MPc
& 2.328 &1.992 &1.988&1.974 &1.975 & 2.021&1.991&1.888
&1.884&1.947&1.988&2.044
 \\[1ex]
\hline
\raisebox{1.0ex}{{\bf m}}&MPc&0.00&0.99&2.00 &3.00 &4.00 &4.80 &2.00 &1.00 &0.00
& 1.00 & 0.00&0.95
 \\[-0.2ex]
\hline
\end{tabular}
\label{table1}
\\
\hspace{0.0cm}{\small $^a$, $^b$, $^c$, $^d$, $^e$ and $^f$ from references \cite{Ki}, \cite{Will}. \cite{Rob}, \cite{Brow}, \cite{Sche} and \cite{Re0}, respectively.}\\
\end{center}
\end{table}

\begin{figure}[h!]
\begin{center}
\includegraphics[width=0.35\textwidth]{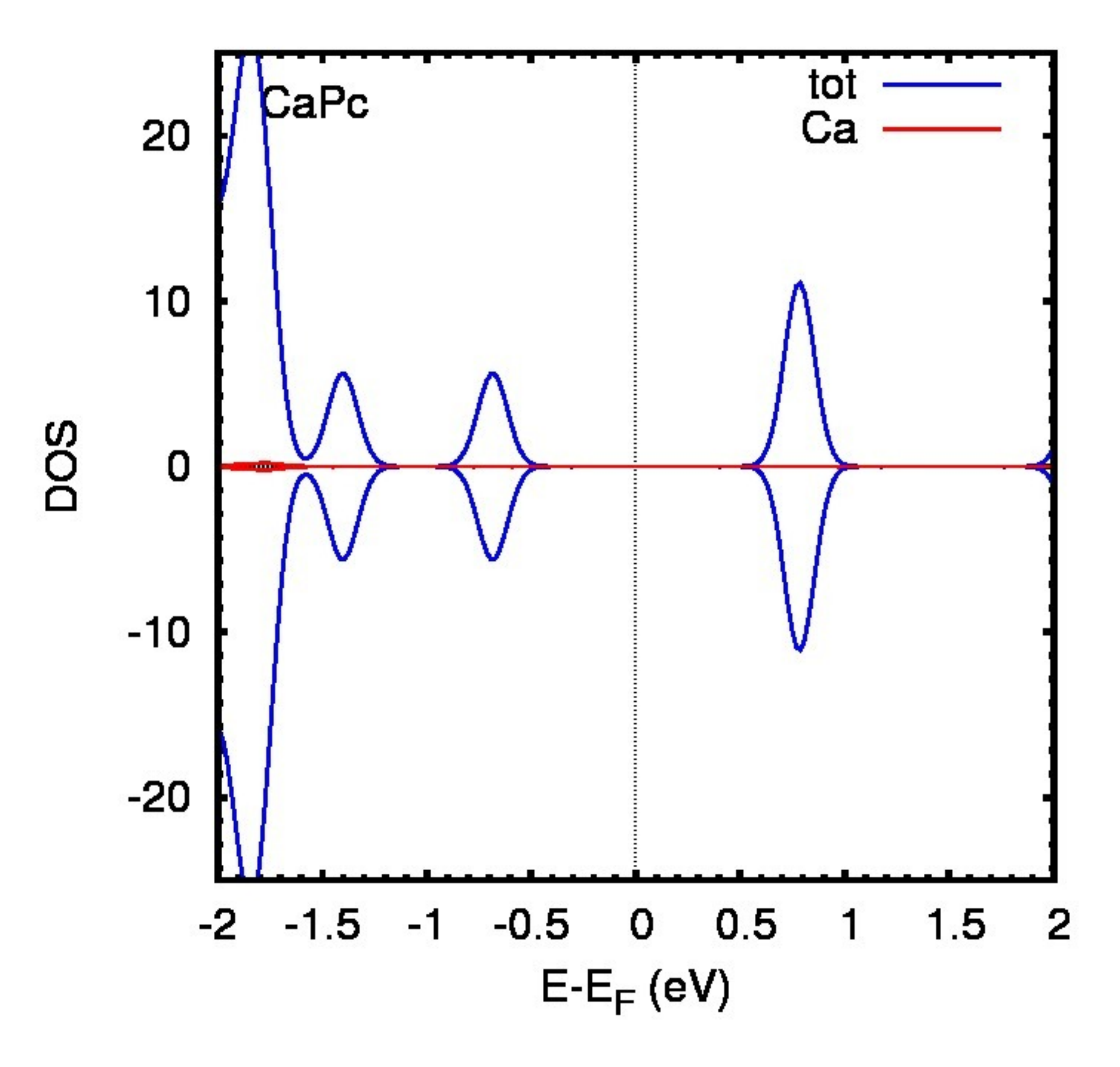}
\includegraphics[width=0.35\textwidth]{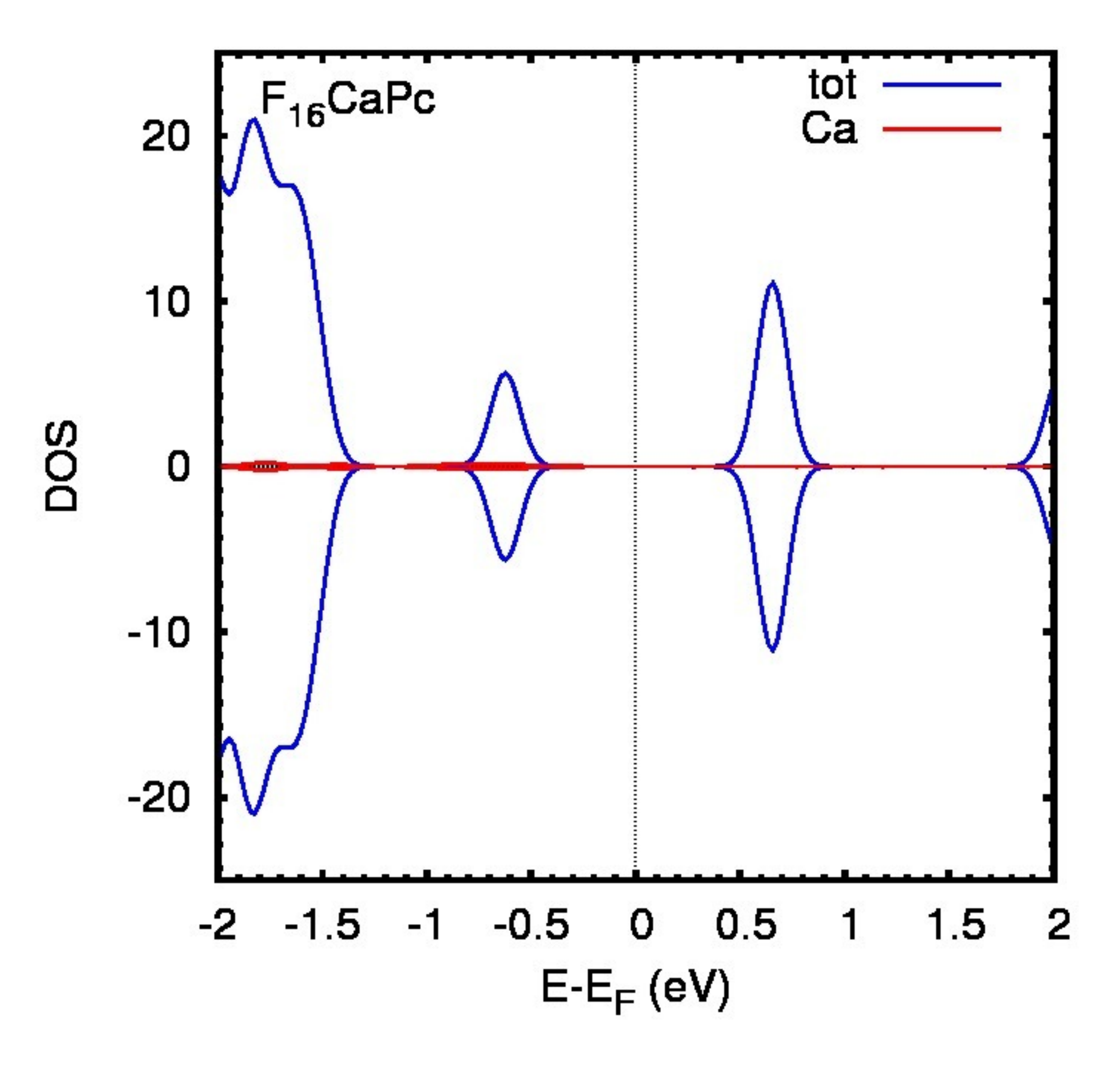}
\end{center}
\begin{center}
\caption{(Color online) Density of states CaPc (left) and for F$_{16}$CaPc (right).}
\label{fig3}
\end{center}
\end{figure}

To complement the analysis of the magnetic properties of the MPc, we discuss their spin resolved electronic structure in the following.
Firstly, we consider the density of states (DOS) of CaPc and F$_{16}$CaPc shown in Fig.\ \ref{fig3}.
Here and in the following plots, the density of states of one of the two spin components is multiplied by $-1$.
Additionally, a level broadening of molecular orbital energies with a spread of 0.1 eV is used to simulate
thermal and anharmonic effects in the spectra.
The energy is shifted with respect to the Fermi level $E_F$, estimated using the Fermi-Dirac function with 
an electronic temperature of 300 K.
The energy gap of CaPc, estimated as the difference between its highest occupied molecular orbital (HOMO) 
and its lowest unoccupied molecular orbital (LUMO), is $1.3$ eV. 
As it can be noted in Fig.\ \ref{fig3}, these orbitals have no metal contribution, and they are localized upon the Pc atoms.
Both peaks are very similar to the fluorinated derivative, the main difference is the decrease of the
energy gap by 0.2 eV. However, the shape of the HOMO--1 (first orbital with an energy below the HOMO energy),
is quite different: it is very broad in F$_{16}$CaPc, in contrast to CaPc where it is rather narrow, 
indicating several energy states close by.

\begin{figure*}
\includegraphics[width=0.3\textwidth]{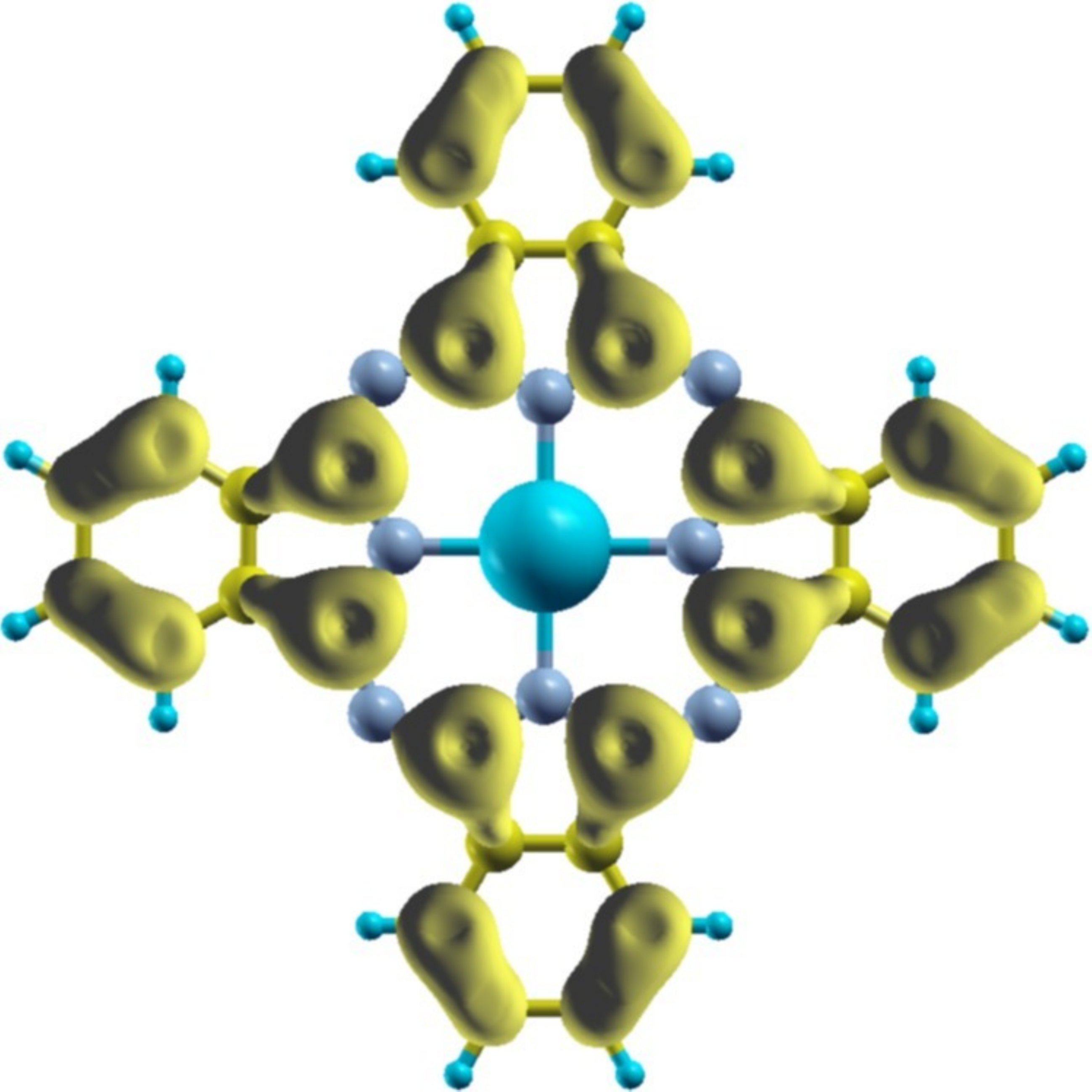}
\hspace{0.4cm}
\includegraphics[width=0.3\textwidth]{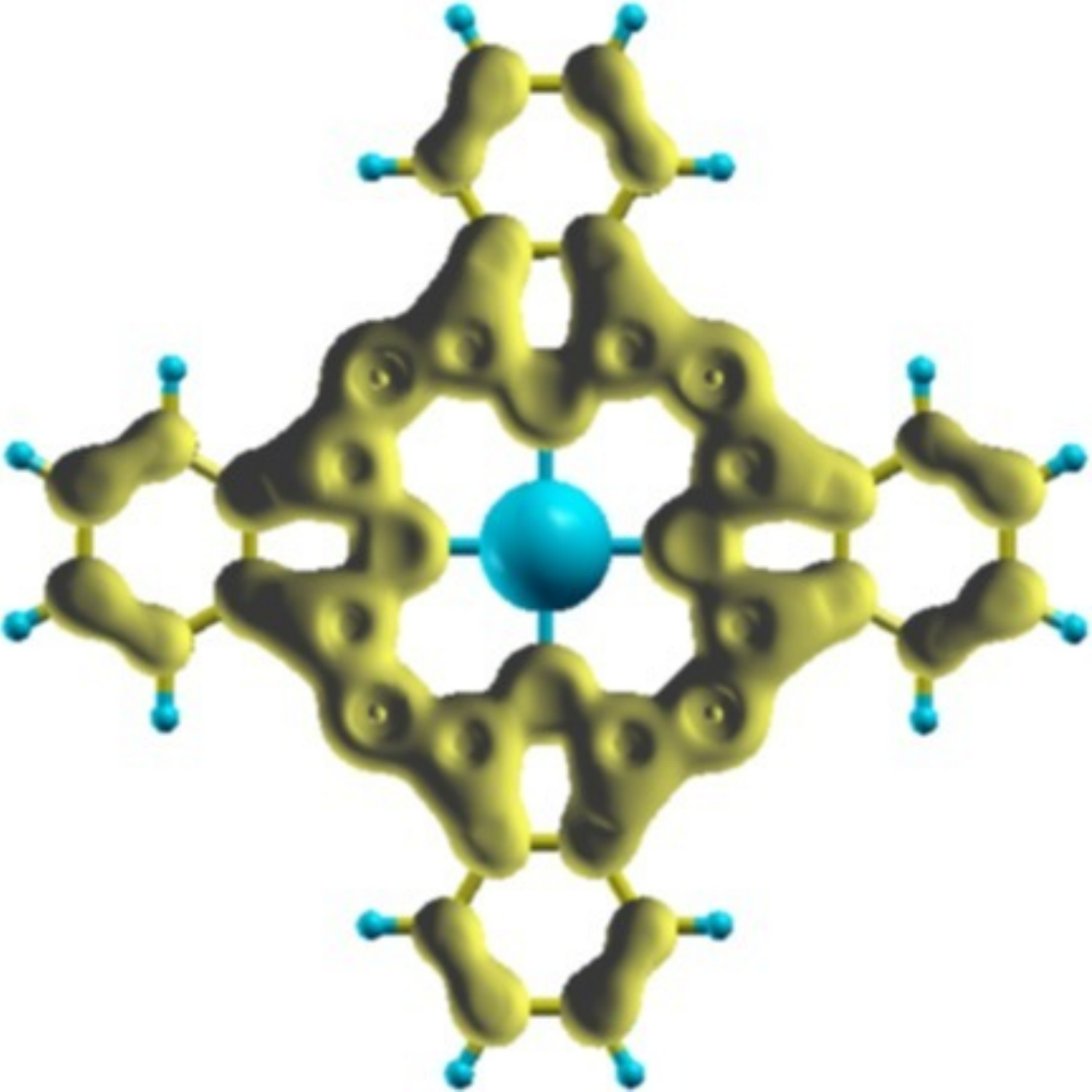}
\caption{(Color online) Charge density isosurface for CaPc: HOMO (left), LUMO (right).}
\label{fig4}
\end{figure*}

To visualize the orbital contributions to the electronic density, we calculate the energy resolved density distribution
$n_{\rm LDOS}({\bf r})=\int_{E_1}^{E_2}{\rm d}E\ N(E,{\bf r})$,
where $N(E,{\bf r})$ is the local density of states. The HOMO and LUMO charge density isosurfaces for
CaPc and F$_{16}$CaPc are shown in Fig.\ \ref{fig4}, confirming that there is no contribution from the metal atom to
these orbitals. This is related to the marked propensity of the alkali metal to loose its valence electrons and to form
ionic bonds, transferring the charge to the MPc backbone.
Dominating contributions in the case of the HOMO are from carbon atoms whereas orbitals localized on nitrogen and
carbon contribute jointly to the LUMO. Therefore this state is more delocalized over the entire inner ring.
The symmetry obtained for the HOMO's density indicates that it is an $a_{1u}$ orbital while the LUMO's density concurs with
an $e_{g}$  orbital type, the relative height of their corresponding DOS peak (see Fig.\ \ref{fig3}) 
indicates that $e_{g}$ is a doubly degenerate level.


The charge density isosurfaces for F$_{16}$CaPc have almost the same shape as CaPc. 
In addition, the fluor atoms have clearly visible contributions.

\begin{figure}[h!]
\begin{center}
\leavevmode
\includegraphics[width=0.3\textwidth,angle=270]{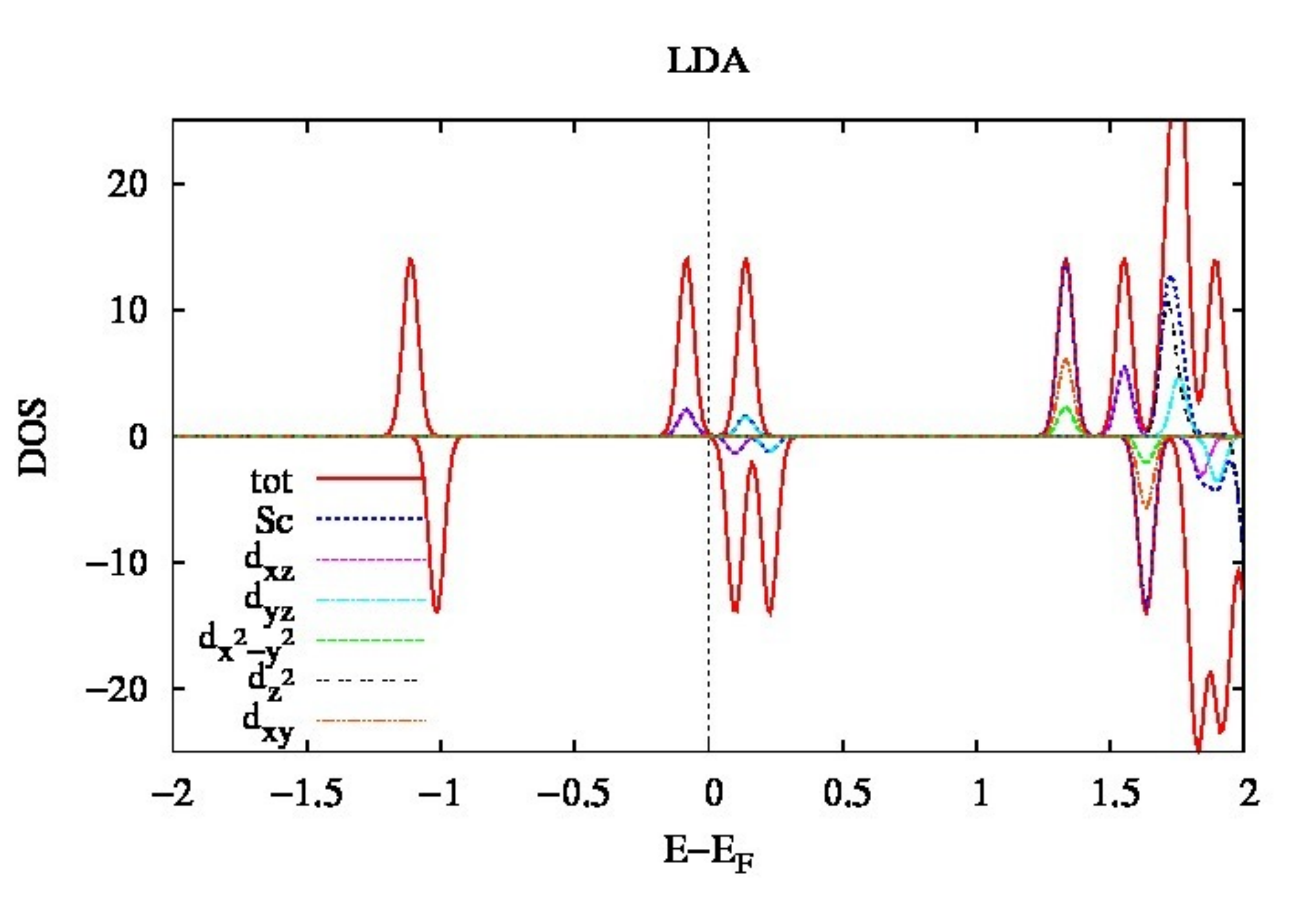}
\includegraphics[width=0.3\textwidth,angle=270]{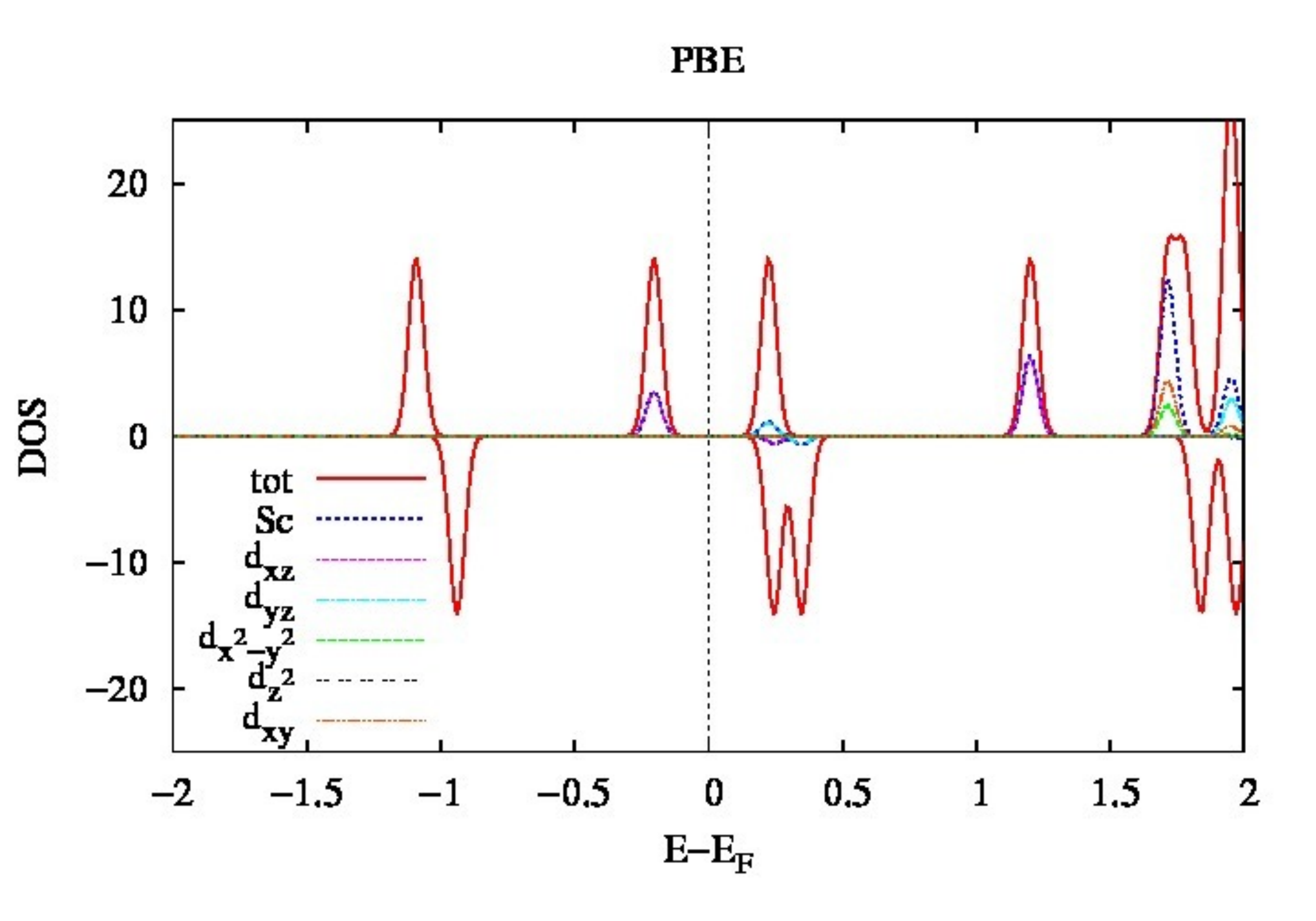}
\end{center}
\begin{center}
\caption{(Color online) Density of states of ScPc with LDA (left) and PBE (right).}
\label{fig5}
\end{center}
\end{figure}

\begin{figure*}
\includegraphics[width=0.3\textwidth]{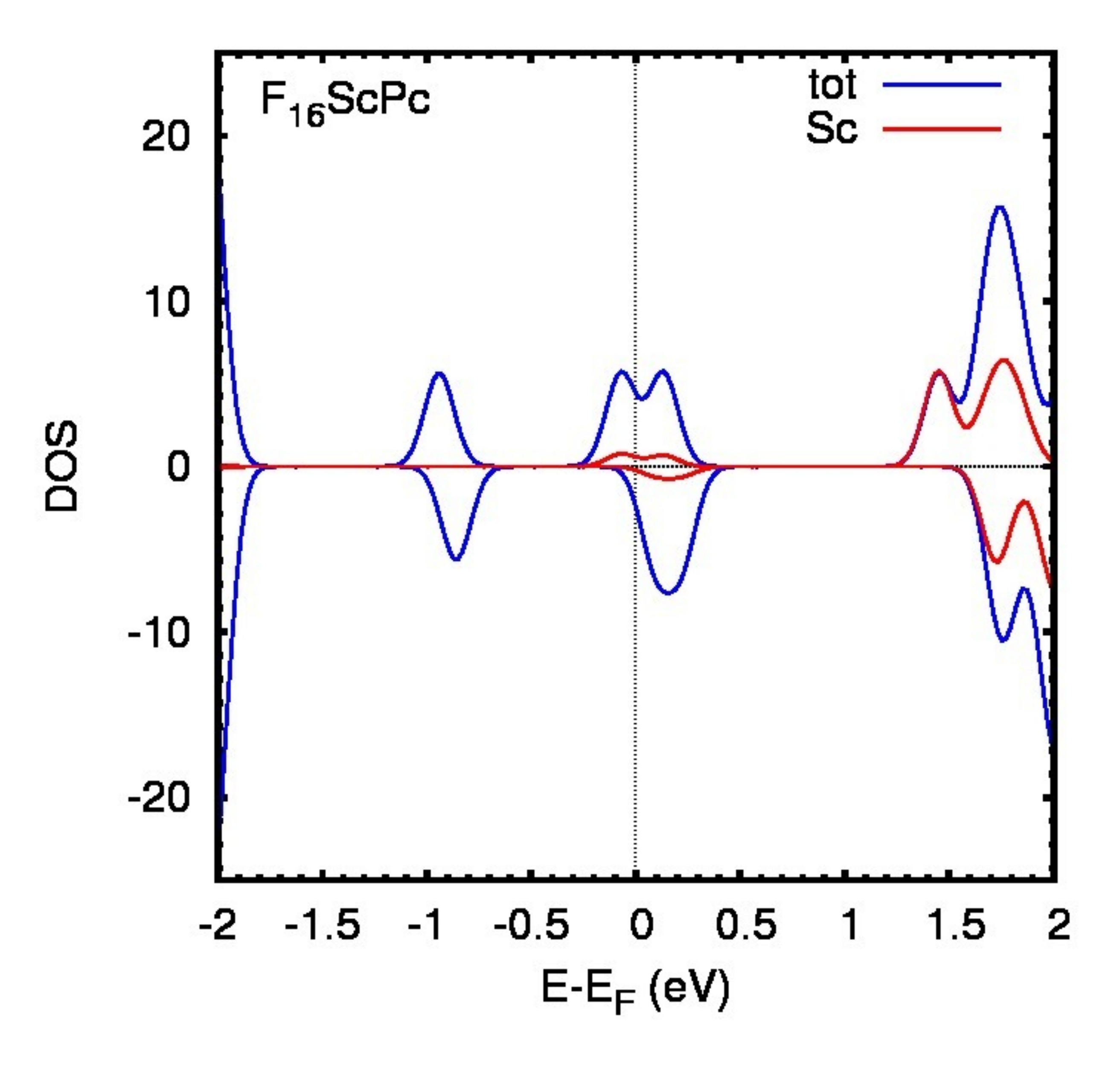}
\caption{(Color online) Density of states of F$_{16}$ScPc with LDA.}
\label{fig6}
\end{figure*}

\begin{figure}[h!]
\begin{center}
\includegraphics[width=0.3\textwidth,angle=270]{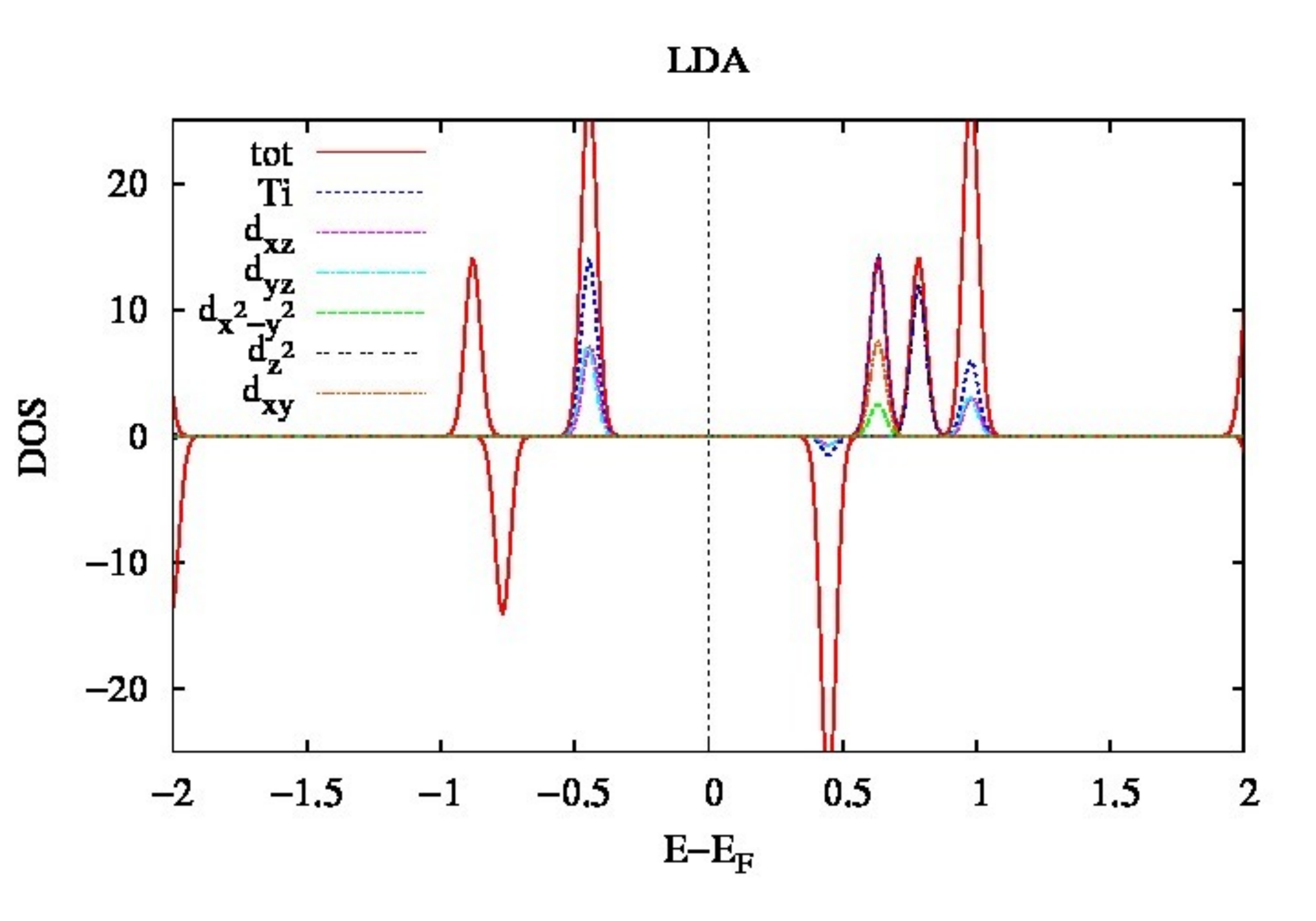}
\includegraphics[width=0.3\textwidth,angle=270]{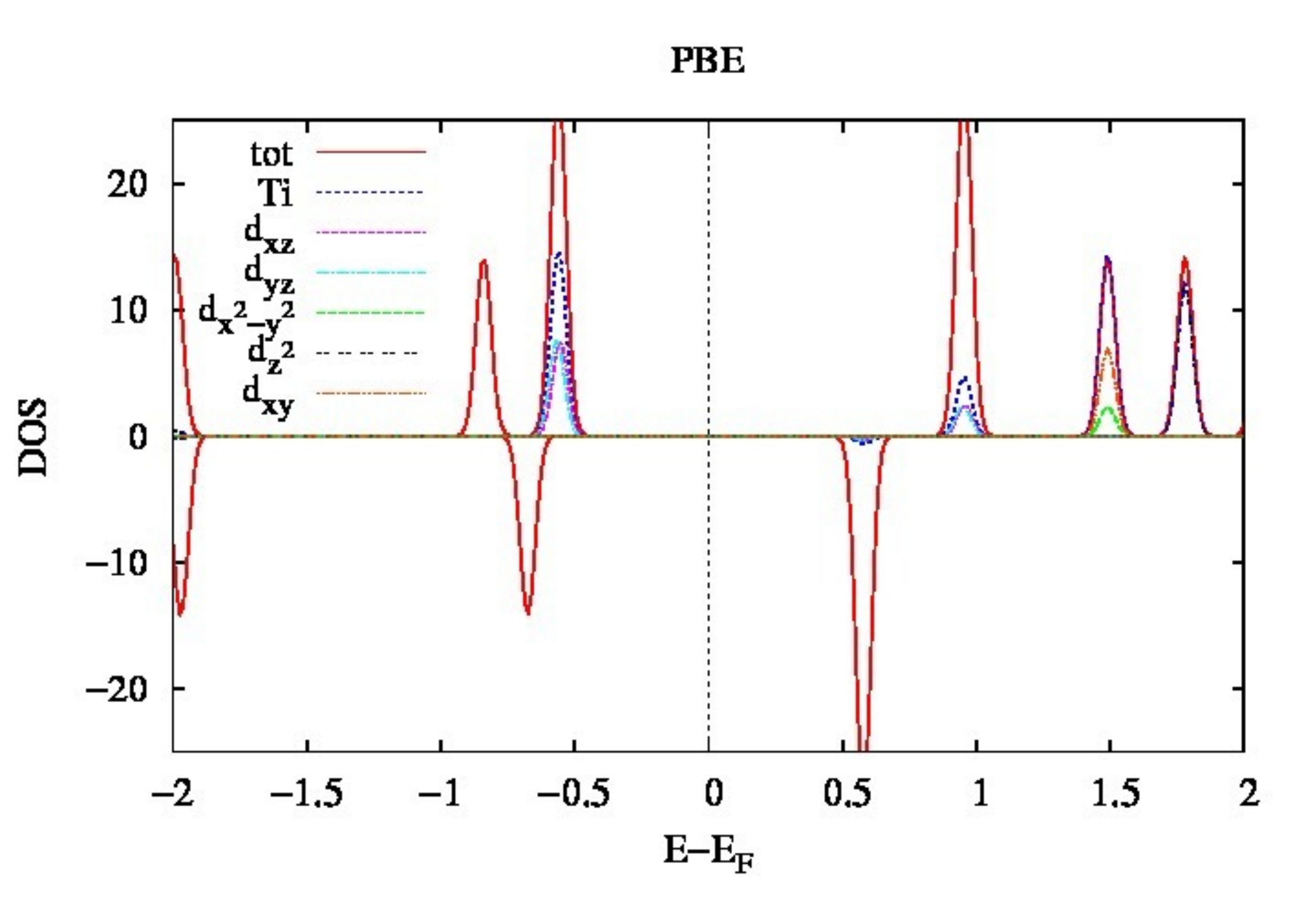}
\end{center}
\begin{center}
\caption{(Color online) Density of states of TiPc with LDA (left) and PBE (right).}
\label{fig7}
\end{center}
\end{figure}

Turning to the $3d$-transition metals, the DOS of ScPc and its fluorinated derivative are shown 
in Fig.\ \ref{fig5} and Fig.\ \ref{fig6}. The states around  $-1.1$ eV are very similar to the HOMO of CaPc.
The Sc $d$ majority states show small contributions just below and above the Fermi level due to the
mixing of the split $d_{xz}$ and $d_{yz}$ 
with the $e_{g}$ ligand orbital which, as mentioned, in CaPc is a LUMO doubly degenerate state.
Metal contributions to unoccupied states also appear above $1.2$ eV, coming from $d_{xy}$, $d_{xz}$, 
$d_{yz}$ and $d_{z^2}$.
The PBE DOS of ScPc indicates that the $d_{xz}$ contribution to the HOMO increases while it decreases 
for unoccupied levels appearing at lower energies (down-shift). 
A wider band gap of 0.3 eV is obtained when PBE is used.
Regarding the electronic structure of TiPc (Fig.\ \ref{fig7}), it can be seen that the metal has more contribution
to the HOMO through its $d_{xz}$ and $d_{yz}$ orbitals than in ScPc.
PBE predicts a slight HOMO down-shift, a marked peak splitting 
of $d_{xy}$ and $d_{z^2}$ metallic contributions and a band gap increase of 0.3 eV. These effects concur with
the expected reduction of the self-interaction error for the semi-local method.

\begin{figure*}
\includegraphics[width=0.3\textwidth]{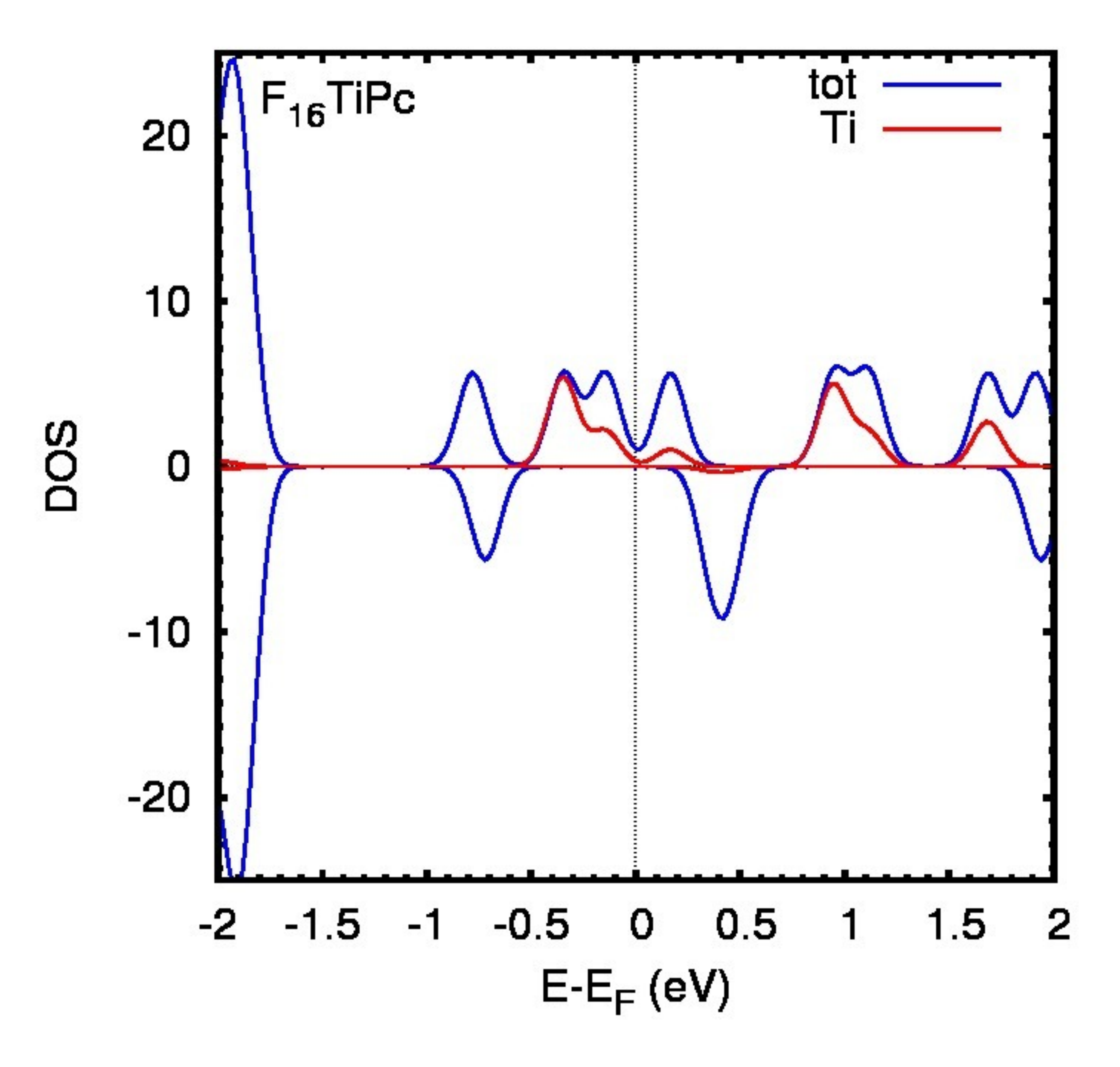}
\caption{(Color online) Density of states of F$_{16}$TiPc with LDA.}
\label{fig8}
\end{figure*}

The electronic structure of F$_{16}$ScPc (Fig.\ \ref{fig6}) indicates that states near the Fermi energy are not influenced 
by fluorination.
In contrast, the main difference between TiPc and F$_{16}$TiPc (Fig.\ \ref{fig8}) is the level splitting of states with 
metallic contribution,
particularly near the Fermi energy. The change in the admixture of Ti and Pc states modifies the majority DOS profile.
Even though fluorination can change the electronic structure, the total magnetic moment is not altered. This is due
to the charge redistribution in the molecule, as can be concluded from peak intensity changes of electronic
structure states close to $E_F$.
The charge deficiency on the metal and its magnetization, estimated as the difference between electronic 
charge densities of opposite spins obtained with the Mulliken population analysis \cite{mulliken}, are reported 
in Tab. \ref{table2}.
The table shows that, with the exception of Ti systems,
the small amount of charge transferred from the metal to Pc is similar for the fluorinated and not 
fluorinated compounds. When the tendency of estimated metal magnetizations is compared to that of the total magnetizations, 
previously shown in Tab. \ref{table1}, it is possible to observe a close interdependence.

As noted in the second row of Tab. \ref{table2}, the estimated metal magnetization difference between 
TiPc and F$_{16}$TiPc 
is $0.23$ $\mu_B$. Along with this increase there is a diminution of 0.11$e$ of the charge transferred from the metal
to the rest of the molecule. This can be understood as a compensation mechanism that explains why both molecules
possess equal total magnetic moments, stressing the ability of these compounds to modify their spin electronic densities
by means of metal--ligand charge redistribution.

\begin{center}
\begin{table}[h]
\caption{Band gap, metal magnetization and metal charge densities of MPc and F$_{16}$
MPc} 
\centering
\begin{tabular}{|c|c|c|c|c|c|c|c|c|c|c|c|c|c|c|}
\hline\hline
 M &&Ca& Sc& Ti& V& Cr& Mn&Fe& Co& Ni& Cu& Zn& Ag
\\ [0.5ex]
\hline
&MPc &1.5  & 0.2 & 0.9 & 0.4 &1.3 & 0.14 & 0.4 & 1.2 & 1.5 & 1.1 &1.5 & 0.4 \\[-1ex]
\raisebox{1.5ex}{Band gap (eV)}&F$_{16}$MPc
& 1.3
 & 0.2 & 0.4 & 0.5 & 1.2 & 0.14 & 0.4 & 1.3 & 1.4 & 1.4 & 1.3 & 0.4 \\[1ex]
\hline
& MPc& 0.00
&0.19&1.47& 2.97 &4.02&4.78 &2.58 &1.31 &0.00 &0.47 & 0.00 & 0.38 \\[-1ex]

\raisebox{1.5ex}{Magnetization ($\mu_B$)} & F$_{16}$MPc& 0.00 
&0.20&1.70& 2.98 &4.03&4.83 &2.57 &1.30 & 0.00 &0.47& 0.00 & 0.38 \\[1ex]
\hline
& MPc& 1.26
&0.30&0.28& 0.15 &0.05 &0.48&0.30 &0.23 &0.14 &0.15 & 0.39 & 0.25 \\[-1ex]

\raisebox{1.5ex}{Charge density (a.u.)} & F$_{16}$MPc& 1.27
&0.28&0.17& 0.16 &0.07&0.48&0.31 &0.24 & 0.15 &0.16 & 0.39 & 0.26 \\[1ex]
\hline
\end{tabular}
\label{table2}
\end{table}

\end{center}

\begin{figure}[h!]
\begin{center}
\leavevmode
\includegraphics[width=0.3\textwidth,angle=270]{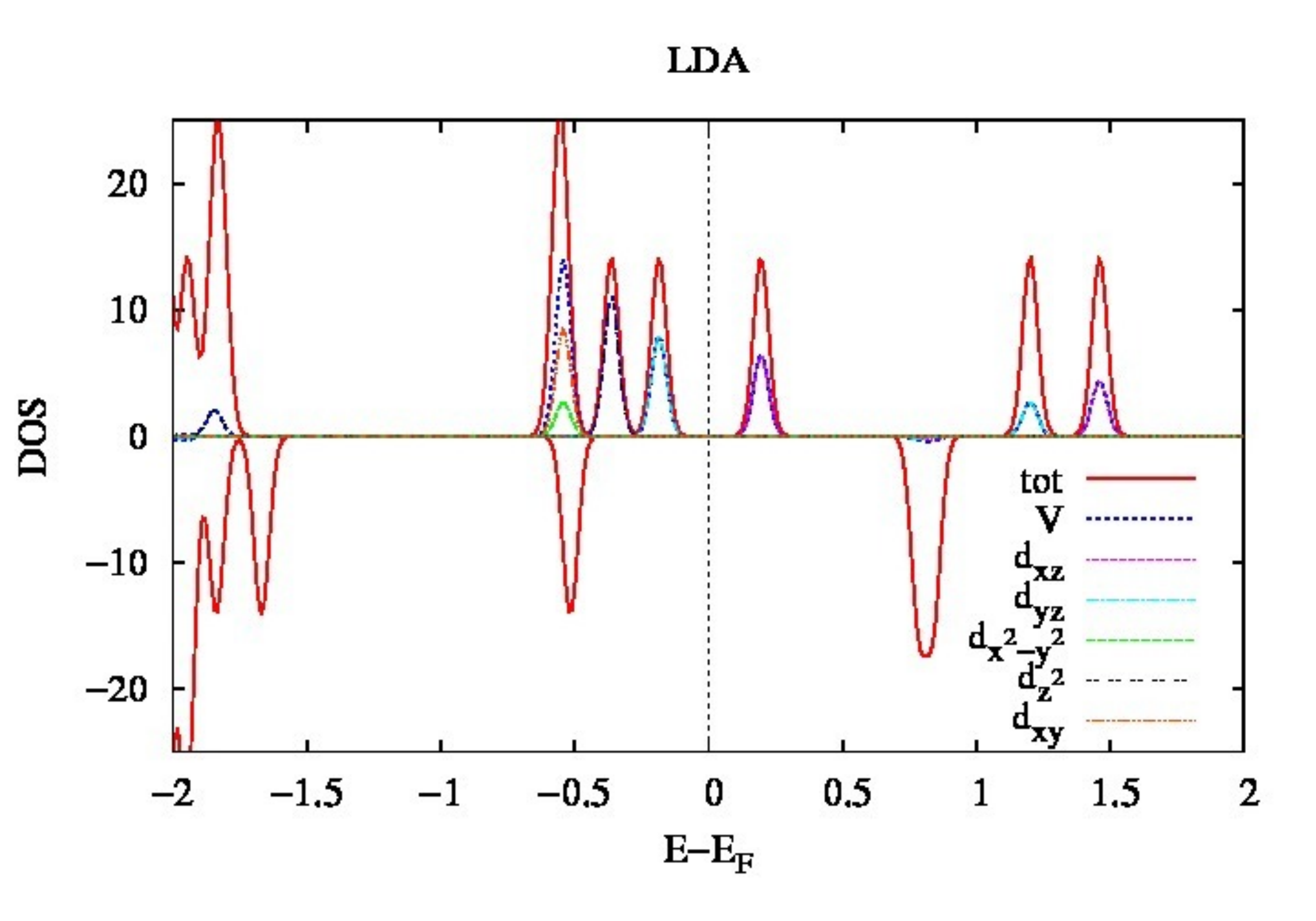}
\includegraphics[width=0.3\textwidth,angle=270]{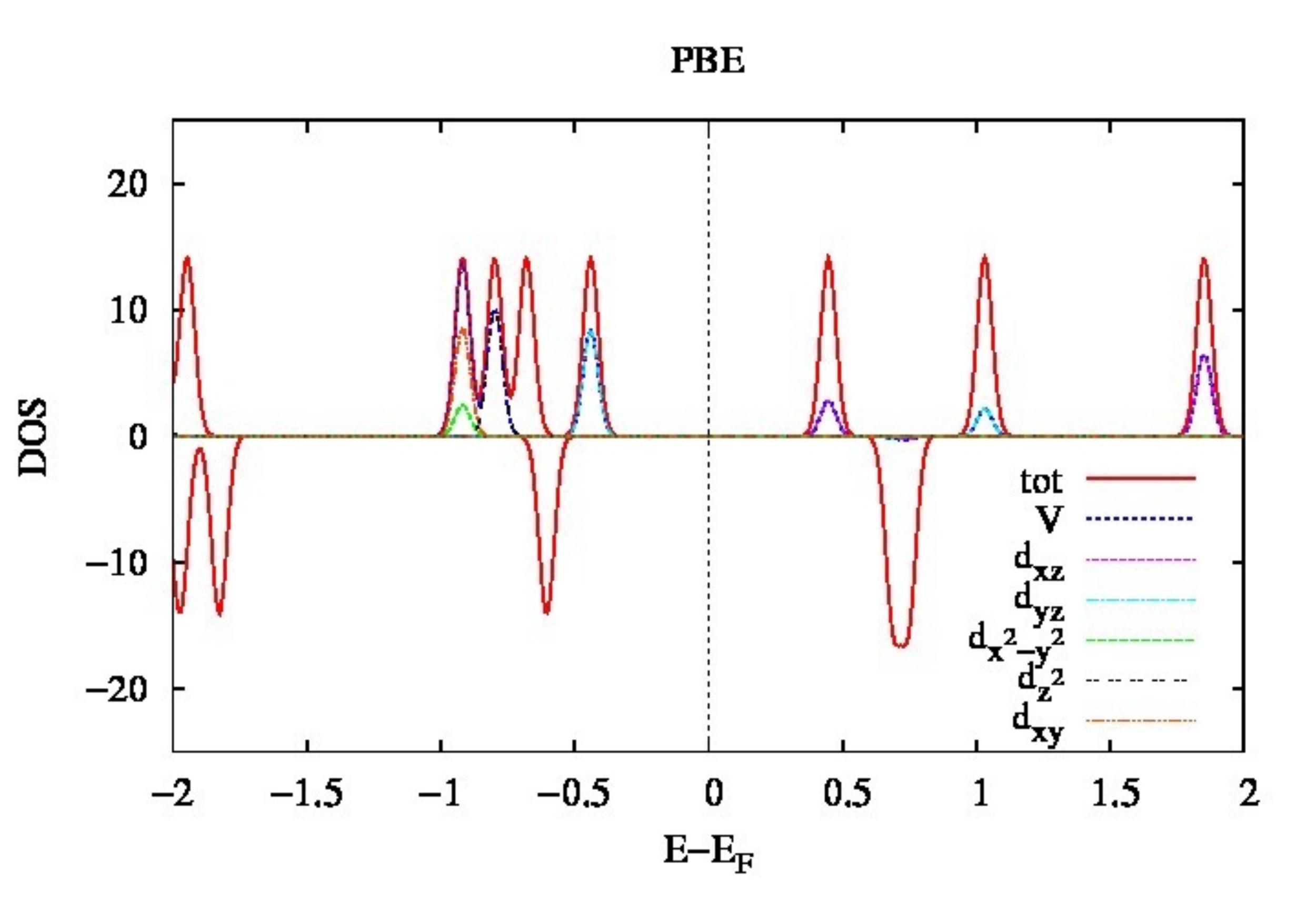}
\end{center}
\begin{center}
\caption{(Color online) Density of states of VPc with LDA (left) and PBE (right).}
\label{fig9}
\end{center}
\end{figure}

Considering the LDA majority DOS of VPc, see Fig.\ \ref{fig9}, we note that new metal orbitals contributing
to the occupied states appear between the HOMO and the $a_{1u}$ molecular orbital. 
They correspond to $d_{z^2}$, $d_{xy}$ and, to a lesser extend, to $d_{x^2-y^2}$. 
The degenerate $d_{xz}$ and $d_{yz}$ metal orbitals in the HOMO of TiPc become split in VPc, and
they compose its HOMO and LUMO. The band gap is reduced to $0.4$ eV.
In contrast PBE predicts this gap to be $0.9$ eV,
displaying a down-shifted HOMO (up-shifted LUMO) with a larger (smaller) metal contribution.
Peak hights of $d_{z^2}$ and $d_{xy}$ are similar to the $d_{yz}$ peak at HOMO, hence they are 
relevant for electron filling along the $d$-row series.
Their energy ordering and particularly the position of  
$d_{xy}$, indicates that the crystalline field effect is not determinant.
In the PBE description the $d_{z^2}$ and $d_{xy}$ metal orbitals appear at lower energies of $a_{1u}$.

\begin{figure}[h!]
\begin{center}
\includegraphics[width=0.3\textwidth, angle=270]{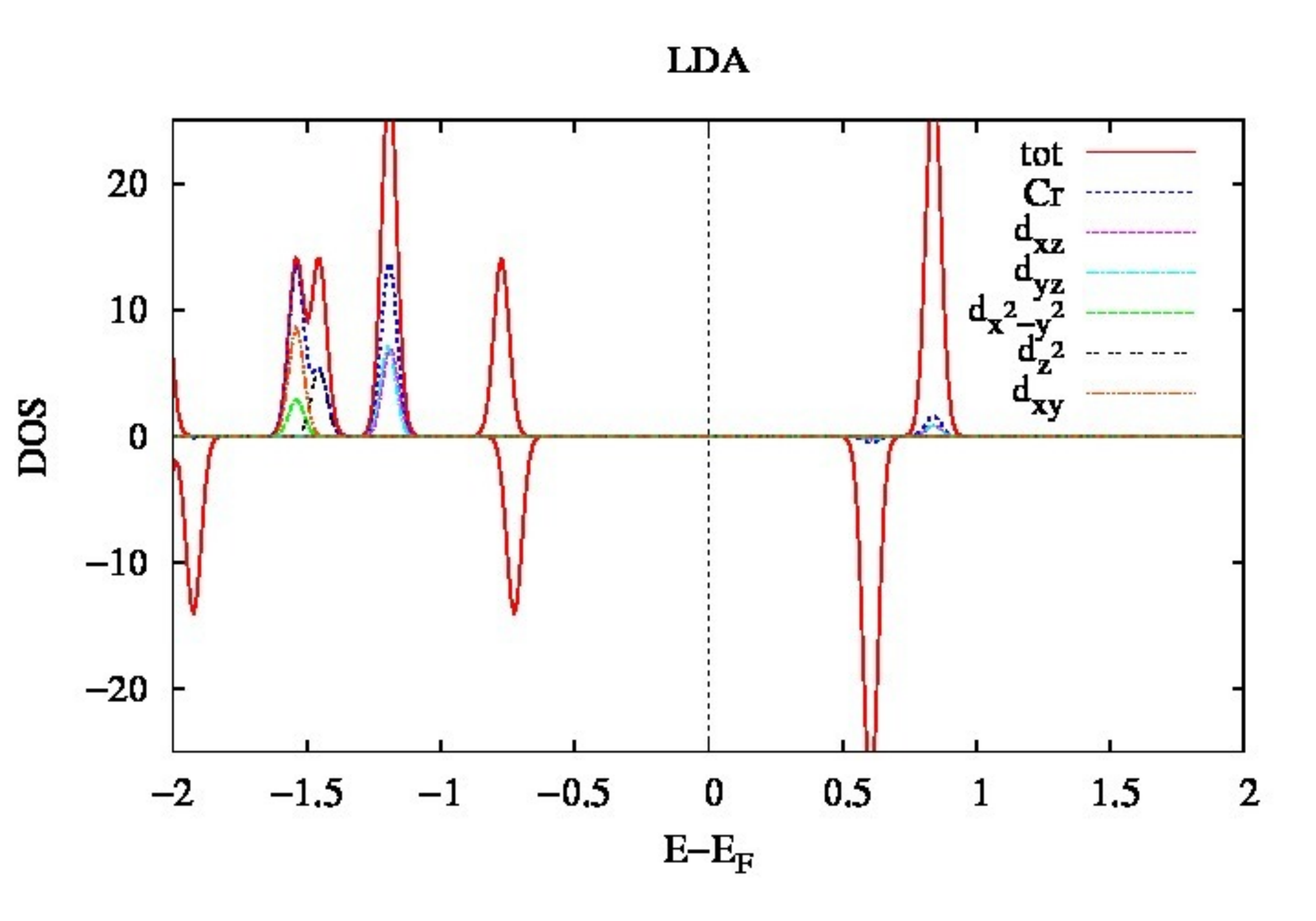}
\includegraphics[width=0.3\textwidth, angle=270]{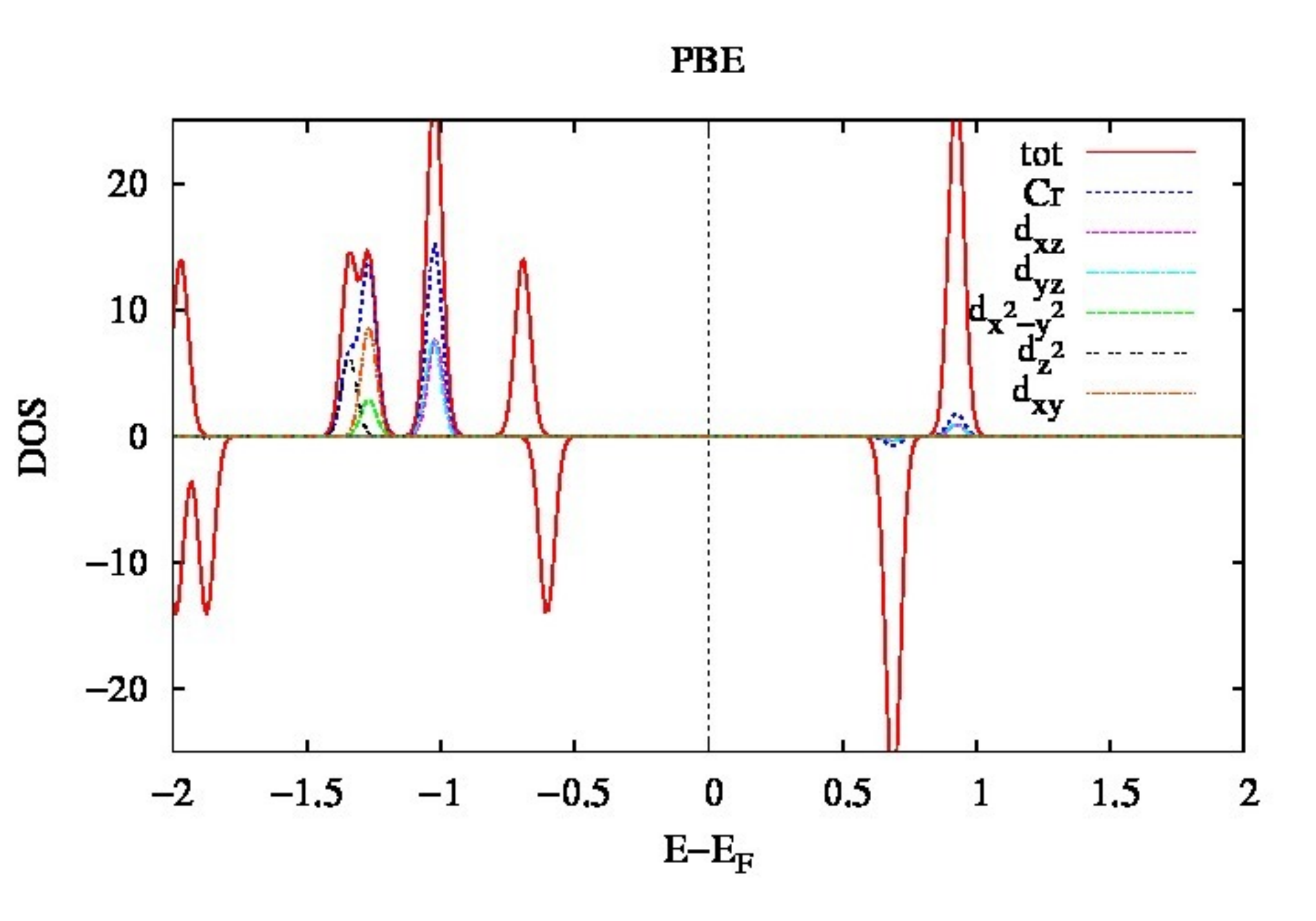}
\end{center}
\begin{center}
\caption{(Color online) Density of states of CrPc with LDA (left) and PBE (right).}
\label{fig10}
\end{center}
\end{figure}

Following along the $3d$ sequence, the additional electron of Cr causes important changes in the electronic DOS, see Fig.\ \ref{fig10}.
Metal contributions to occupied levels appear below the HOMO. The $d_{xz}$ and $d_{yz}$ orbitals which are split in VPc
become degenerate as in TiPc. For the first time within this MPc series the closest orbitals to the Fermi energy 
show larger energy differences for the majority states than for the minority states.
It can be seen that there are no metal contributions to
HOMO and LUMO, they principally localize upon the inner ring. HOMO--1 and HOMO--2 (first and second orbital below 
the HOMO energy) show only a small down-shift with PBE but the band gap remains almost the same.

\begin{figure}[h!]
\begin{center}
\includegraphics[width=0.3\textwidth,angle=270]{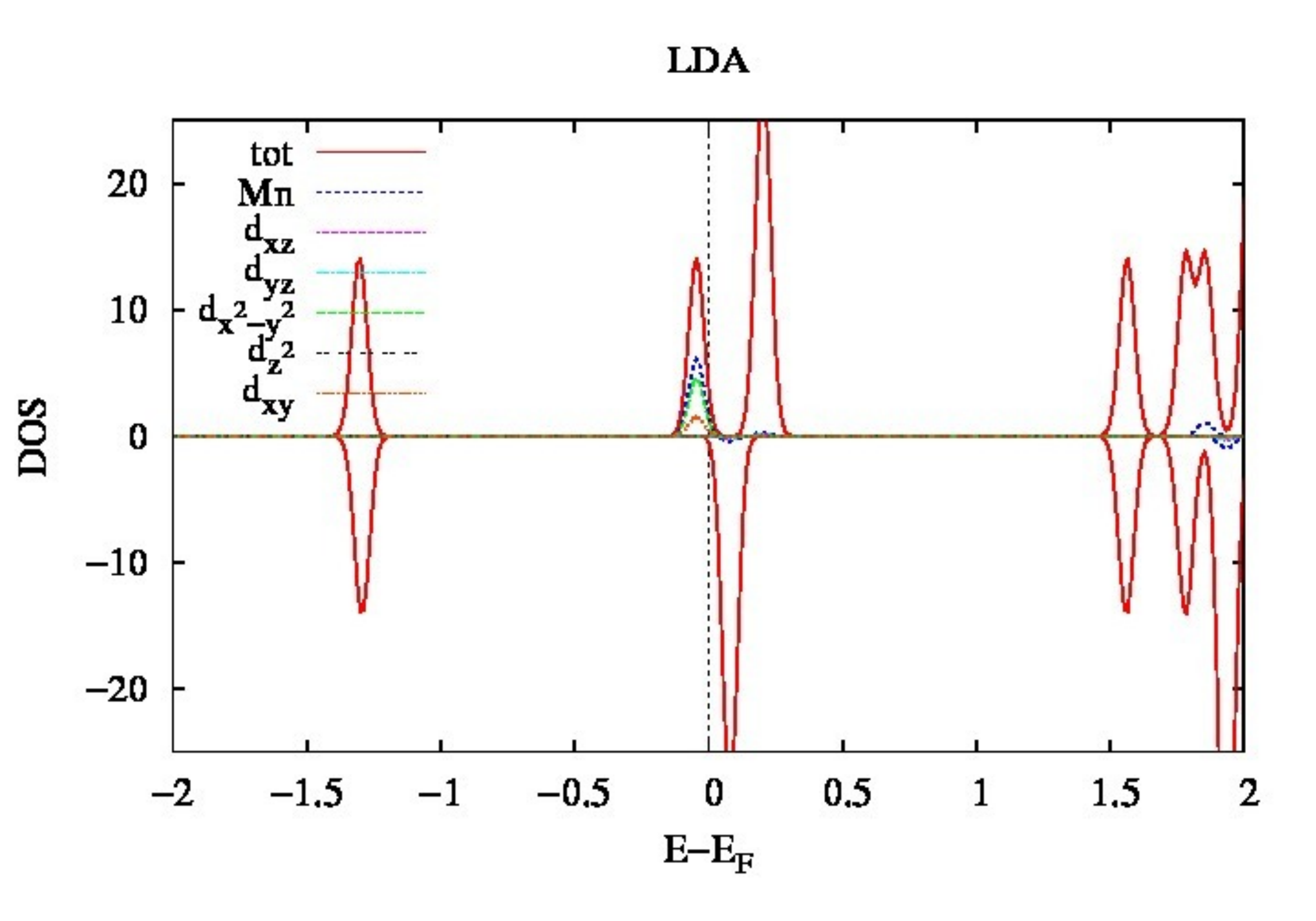}
\includegraphics[width=0.3\textwidth,angle=270]{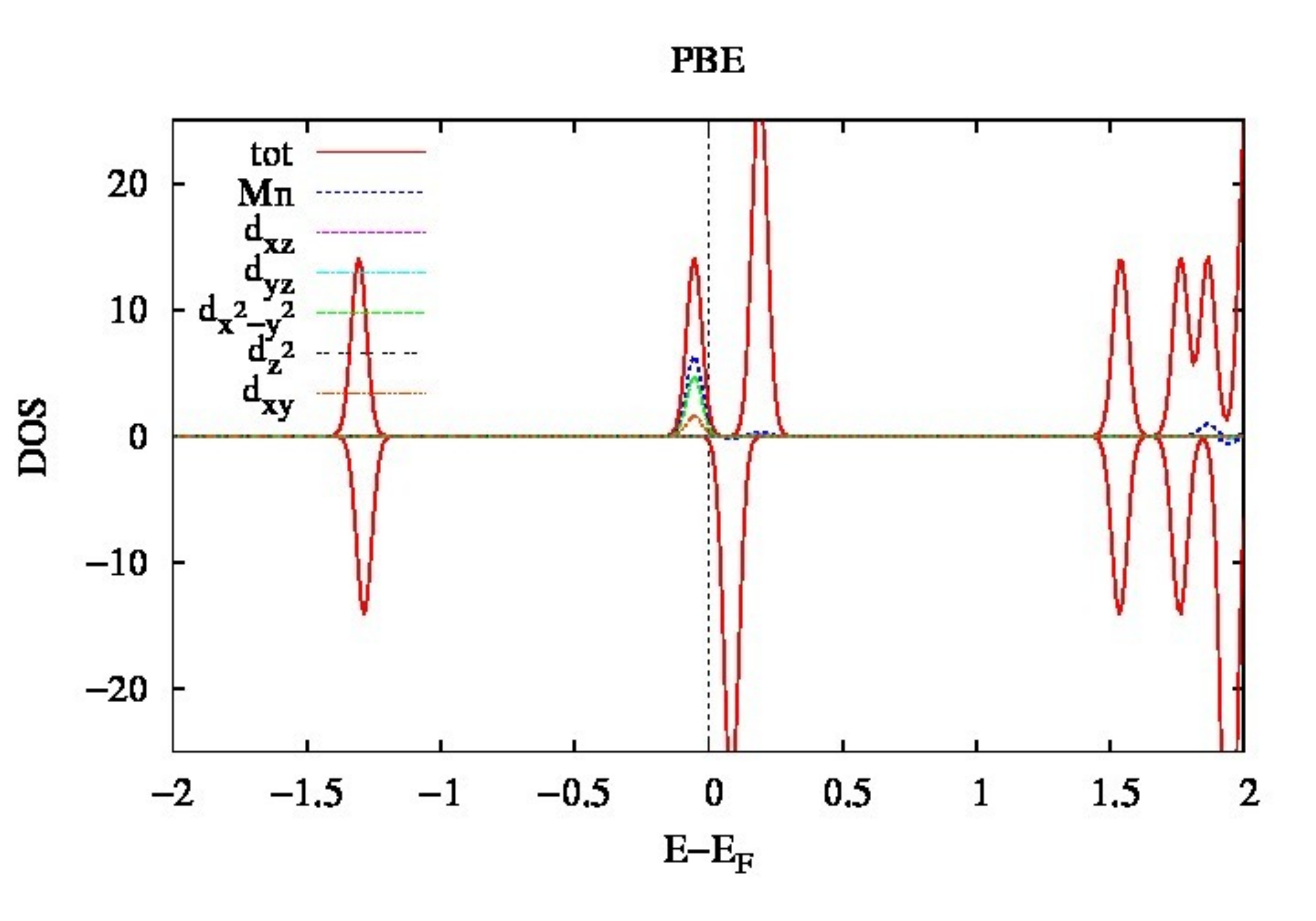}
\end{center}
\begin{center}
\caption{(Color online) Density of states of MnPc with LDA (left) and PBE (right).}
\label{fig11}
\end{center}
\end{figure}

\begin{figure}[h!]
\begin{center}
\includegraphics[width=0.3\textwidth,angle=270]{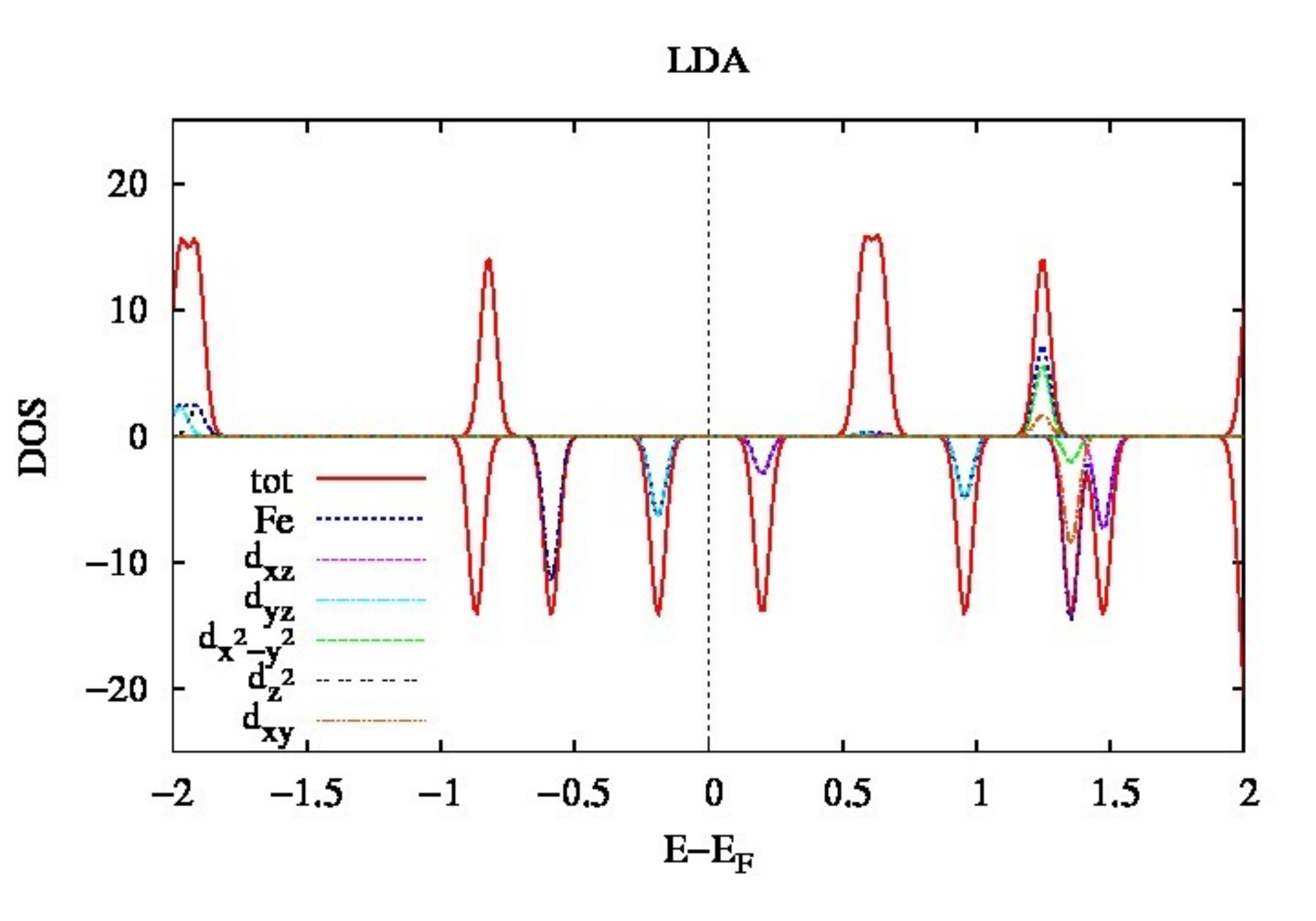}
\includegraphics[width=0.3\textwidth,angle=270]{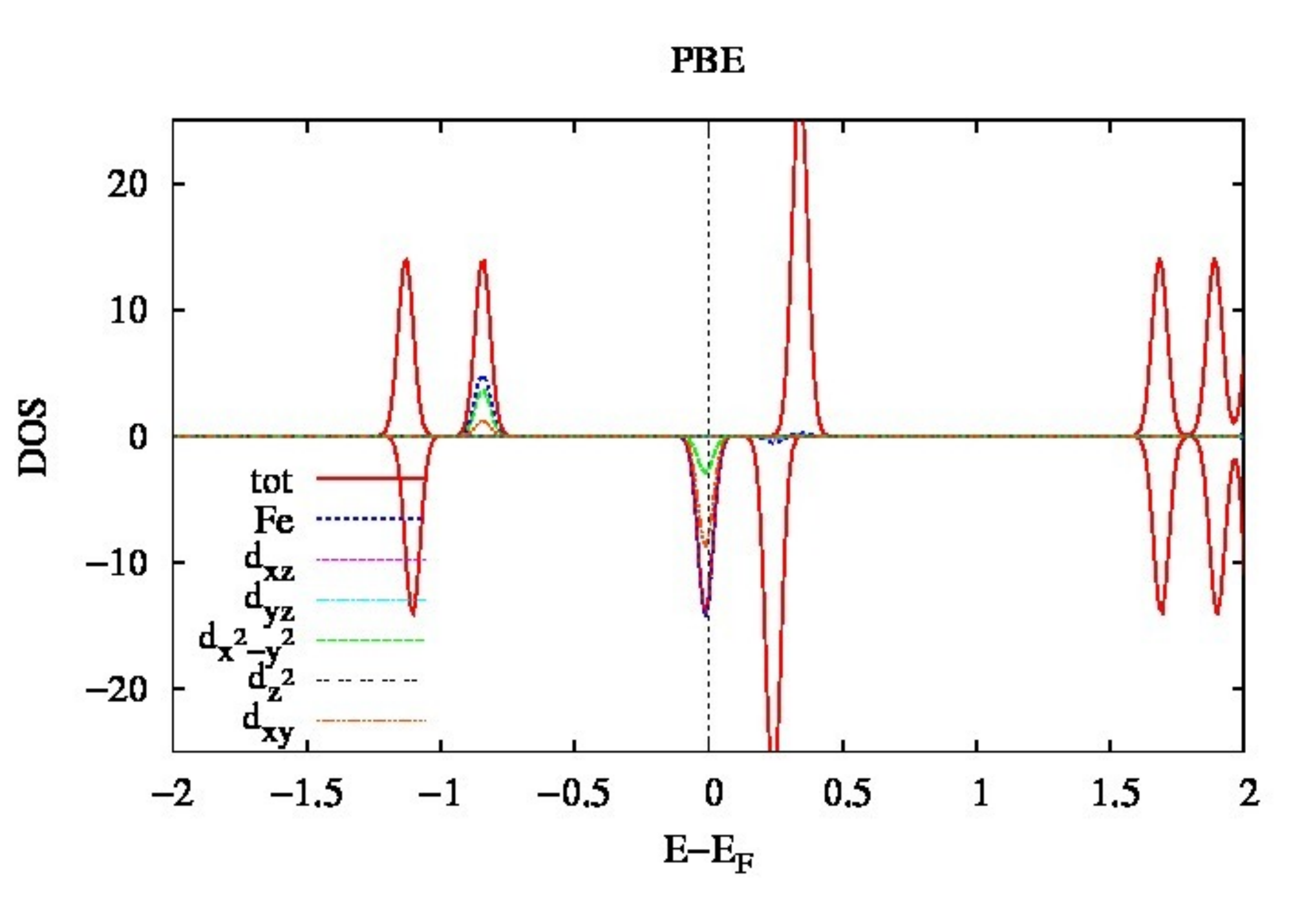}
\end{center}
\begin{center}
\caption{(Color online) Density of states of FePc with LDA (left) and PBE (right).}
\label{fig12}
\end{center}
\end{figure}

The densities of states of MnPc and FePc are shown in Fig.\ \ref{fig11} and Fig.\ \ref{fig12}, respectively. 
Inside the $-2.0$ ... $2.0$ eV window around the Fermi energy, the only Mn non-negligible contribution 
corresponds to the $d_{x^2-y^2}$ ($b_{1g}$) to the HOMO.
Other metal contributions to occupied states, previously seen in VPc and CrPc, appear at energies
about $0.5$ eV below the HOMO--1 (a$_{1u}$).
The LUMO displays only ligand contribution, then the closest orbitals to the Fermi energy have 
a metal--ligand character with a narrow band gap of $0.14$ eV (within LDA), and $0.18$ eV (within PBE).
These values are close to the PBE result, $0.21$ eV, of Stradi et al. \cite{stradi11}  
who used different
double-zeta (6-31G(d,p)) and triple-zeta basis sets for localized Gaussian-type orbitals (GTO), 
as well as plane-wave basis sets.
Their peak decomposition analysis of DOS indicates
that the metal contributes to HOMO and LUMO through $d_{yz}$ and $d_{xz}$; 
this result holds for every method they tested although it significantly reduces for LUMO with PBE0 and HSE06.
Different functionals (PBE, B3LYP, PBEh and M06) have predicted either of two configurations if symmetry
is enforced, $a_{1g}\downarrow$(HOMO) $e_{g}\downarrow$(LUMO) and $b_{2g}\downarrow$(HOMO) $e_{g}\downarrow$(LUMO) \cite{Marom09b}.
On the other hand, when symmetry restrictions are not used either of two configurations are obtained, 
$e_{g}\downarrow$(HOMO) $e_{g}\uparrow$(LUMO) and $e_{g}\downarrow$(HOMO) $e_{g}\downarrow$(LUMO), as reported in
the same work.
Qualitatively, our result agrees with those configurations with a metal (ligand) contribution to the HOMO (LUMO) state.
Both theoretical works point out that the calculated eigenvalue spectra of isolated MnPc appears compressed 
(into a smaller energy window) 
by about 20 \% with respect to the reported data of the UPS experiment for thin and thick films
deposited on metal surfaces. 
A more direct comparison is not possible because data for the gas phase are not available, and our calculations are
basically for isolated molecules.

In FePc, the addition of one electron initiates the second half of the $d$-shell, where now the atomic orbitals can have 
two electrons, bringing about more possibilities to accommodate electrons and increasing the electron correlation due to
the electron pairing.
The HOMO and LUMO can be identified as $e_{g}$ orbitals with $d_{yz}$ 
and $d_{xz}$ metal contributions, respectively, 
when using LDA, and $d_{x^2-y^2}$ metal orbital and $e_{g}$ with PBE. Regarding the metal (ligand) character of HOMO (LUMO), 
our PBE and LDA result concurs with recent calculations \cite{brena12,bruder10,nakamura12,mugarza12} but deviates from 
another prediction \cite{Marom09b}.
Furthermore, several theoretical works devoted to the interpretation of experimental results differ in the assessment of metal or ligand character of HOMO and LUMO. 
The analysis of electron energy-loss spectral data for FePc films \cite{epjb2010} strongly suggests that partially 
occupied states $e_{g}$, with $d_{yz}$ and $d_{yz}$ metal contributions, lie in between $a_{1u}$ and $e_{g}$ ligand states, 
manifesting metal participation for HOMO with an energy difference to LUMO ($e_{g}$ orbital on the ligand) of $0.45$ eV,
very close to our LDA band gap, $0.4$ eV, and PBE, $0.5$ eV, estimates.
The occurrence of metal orbitals close to the Fermi level, particularly for HOMO, has also been concluded 
from the analysis and 
interpretation of photoemission and X-ray absorption spectroscopy measurements of FePc on Ag(111) \cite{petraki12}.
The metal--ligand character obtained with LDA for HOMO--LUMO is used to describe the charge transfer between FePc and Au(111) 
 \cite{minamitani12} by means of the LDA+U method. 
Analogously the GGA+U method is employed to reproduce charge reorganization of FePc adsorbed on Ag(100) deduced from
STM spectrospcopy \cite{mugarza12}.
Apart from the +U correction our results differ in a qualitative manner because we find $d_{z^2}$ states 
close the Fermi level but not
$d_{xy}$, instead we get $d_{x^2-y^2}$. This may be due to a larger degree of metal--ligand hybridization for $d_{xy}$.
In contrast a ligand--ligand character is suggested to account for photoelectron spectroscopy experiments
which help to determine $U_{\rm eff}$ within the GGA+U correction \cite{brena12}.
Nevertheless, recent total-energy density functional calculations together with density-matrix constraints indicate that 
isolated and columnar stacking FePc molecules ($\alpha$-FePc) ground states have a substantial $d$ metal orbital 
contribution to HOMO \cite{nakamura12}.

\begin{figure}[h!]
\begin{center}
\includegraphics[width=0.3\textwidth,angle=270]{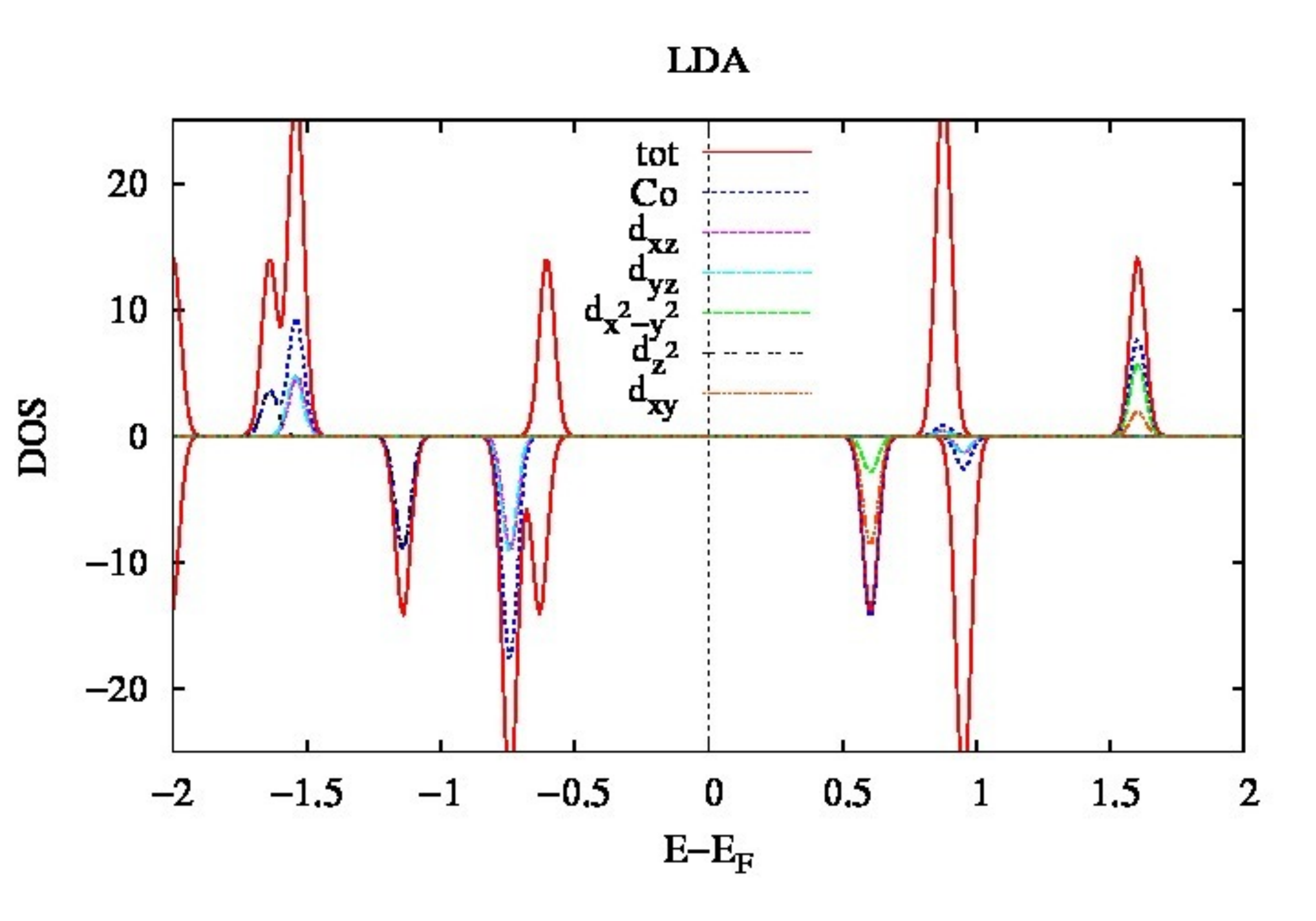}
\includegraphics[width=0.3\textwidth,angle=270]{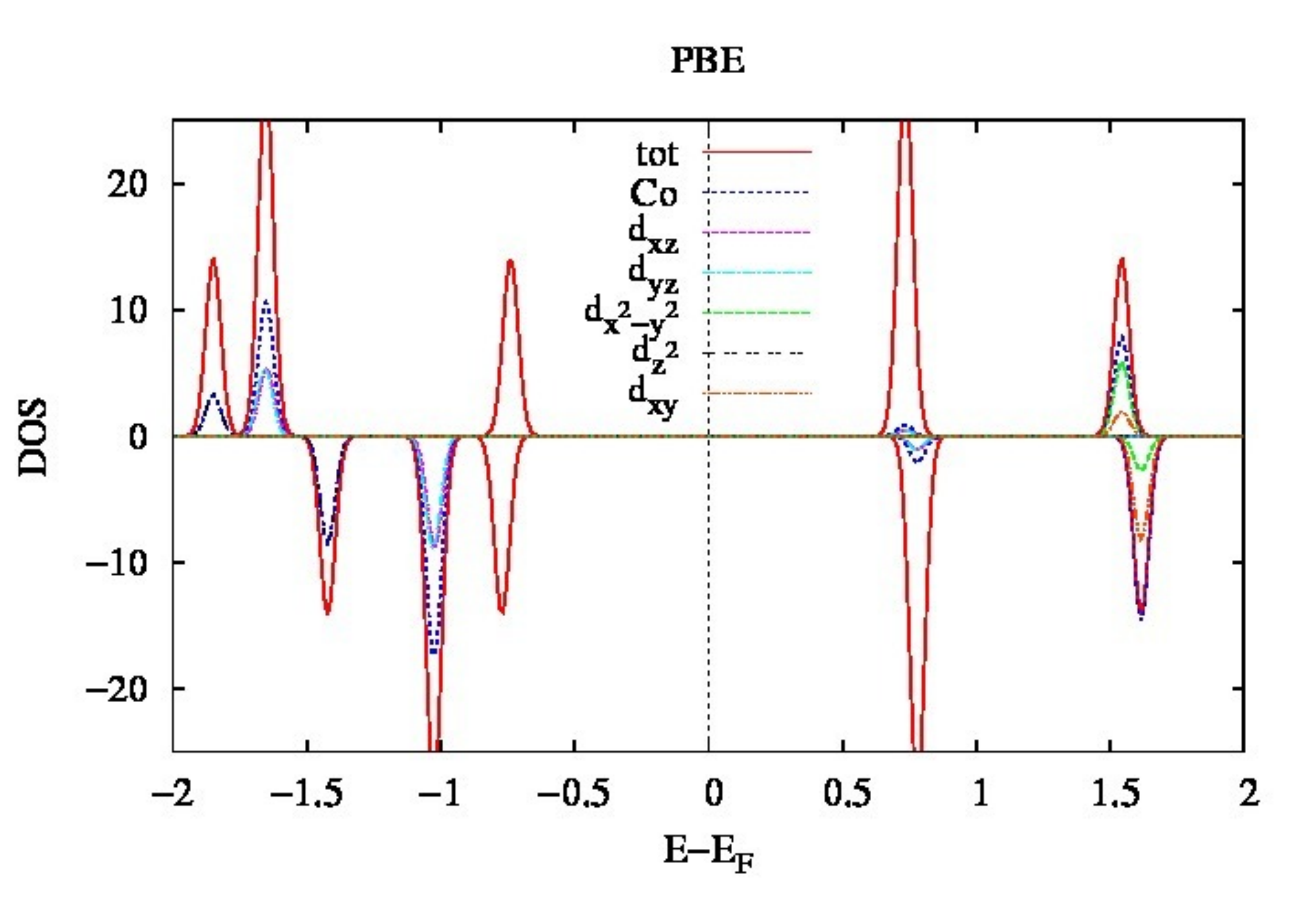}
\end{center}
\begin{center}
\caption{(Color online) Density of states of CoPc with LDA (left) and PBE (right).}
\label{fig13}
\end{center}
\end{figure}

\begin{figure*}
\includegraphics[width=0.3\textwidth]{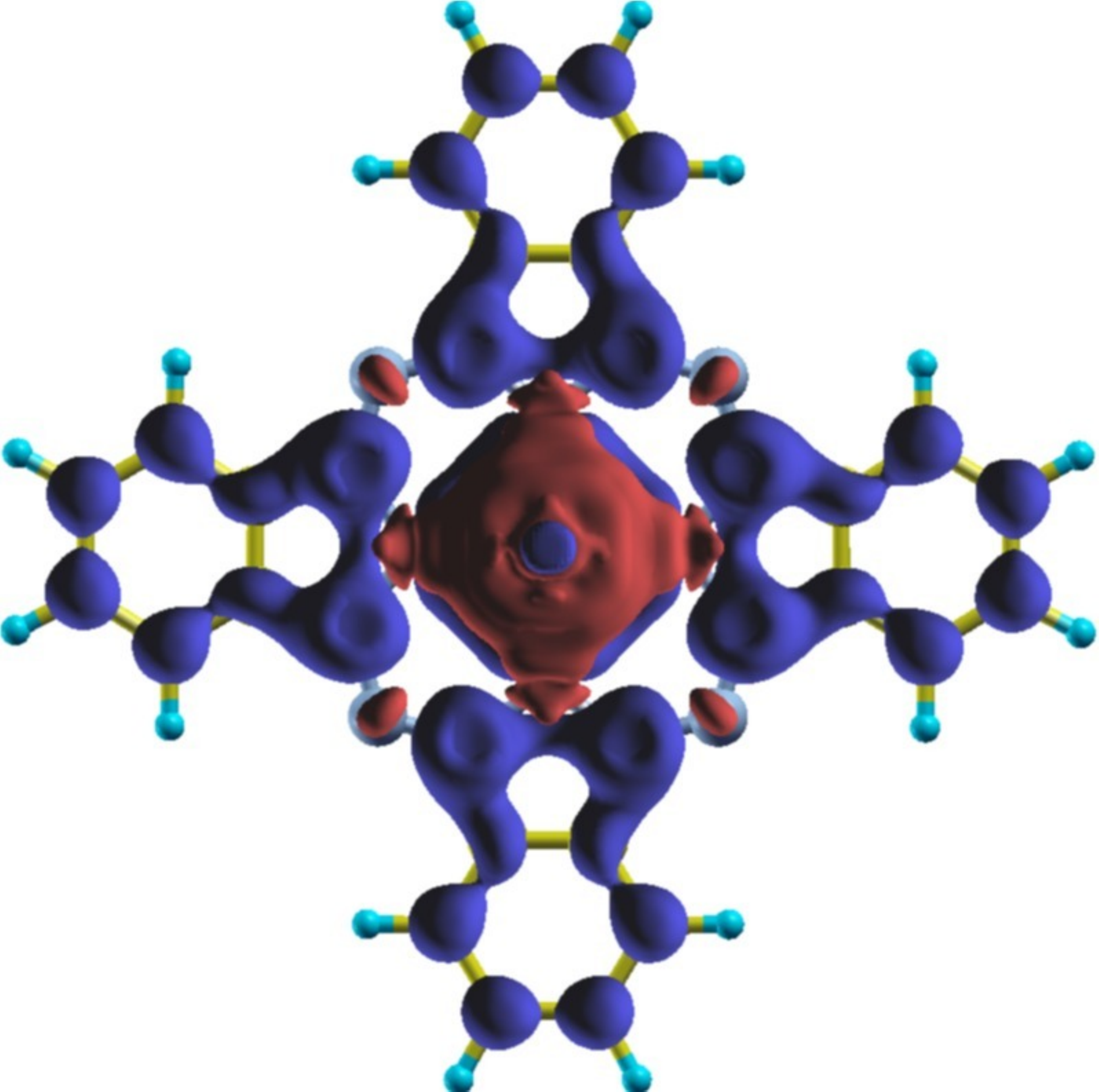}
\caption{(Color online) Charge density isosurface for occupied states, red minority-spin and blue majority-spin electron density.}
\label{fig14}
\end{figure*}

Figure \ref{fig13} displays the density of states for CoPc.
The LDA band gap, $1.2$ eV, is larger than the previously reported result, 1.0 eV \cite{Re3}, using LDA and GGA+vdW.
The band gap obtained with GGA, $1.45$ eV, is the same as the one predicted for a CoPc monolayer \cite{Bi4} 
by the full-potential linearized
augmented plane wave (FLAPW) method, and very close to the GGA+U result of $1.4$ eV \cite{mugarza12}, 
but smaller than the calculated 
value of $1.96$ eV \cite{Me} using the Vosko-Wilk-Nusair (VWN) potential \cite{R0} including exchange 
and correlation gradient corrections.
In the majority component, LDA and PBE indicate that HOMO and LUMO are distributed over the Pc atoms. 
From the charge density isosurface (Fig.~\ref{fig14}) of the occupied Co states from $-2$ eV to $E_{F}$, we see that 
the contribution to the majority density principally derives from Co and N states while carbon atoms contribute to minority states.
In regard to the LDA DOS, we find that the metal-localized occupied $d_{z^2}$ orbital down-shifts in the GGA description, 
on the contrary the LUMO, $d_{xy}$, up-shifts and the $2e_g$ orbital becomes LUMO. 
The ligand ($a_{1g}$) -- ligand ($2e_g$) character 
of HOMO -- LUMO, differs from the metal ($d_{z^2}$) -- ligand \cite{Bi4}, ligand -- metal \cite{zhang11}, 
ligand ($1e_g$) -- ligand ($2e_g$) \cite{Me} character of other works, 
but agrees with recent theoretical results \cite{mugarza12} in the gas phase, carried out to account for STM experimental observations.

\begin{figure}[h!]
\begin{center}
\includegraphics[width=0.3\textwidth,angle=270]{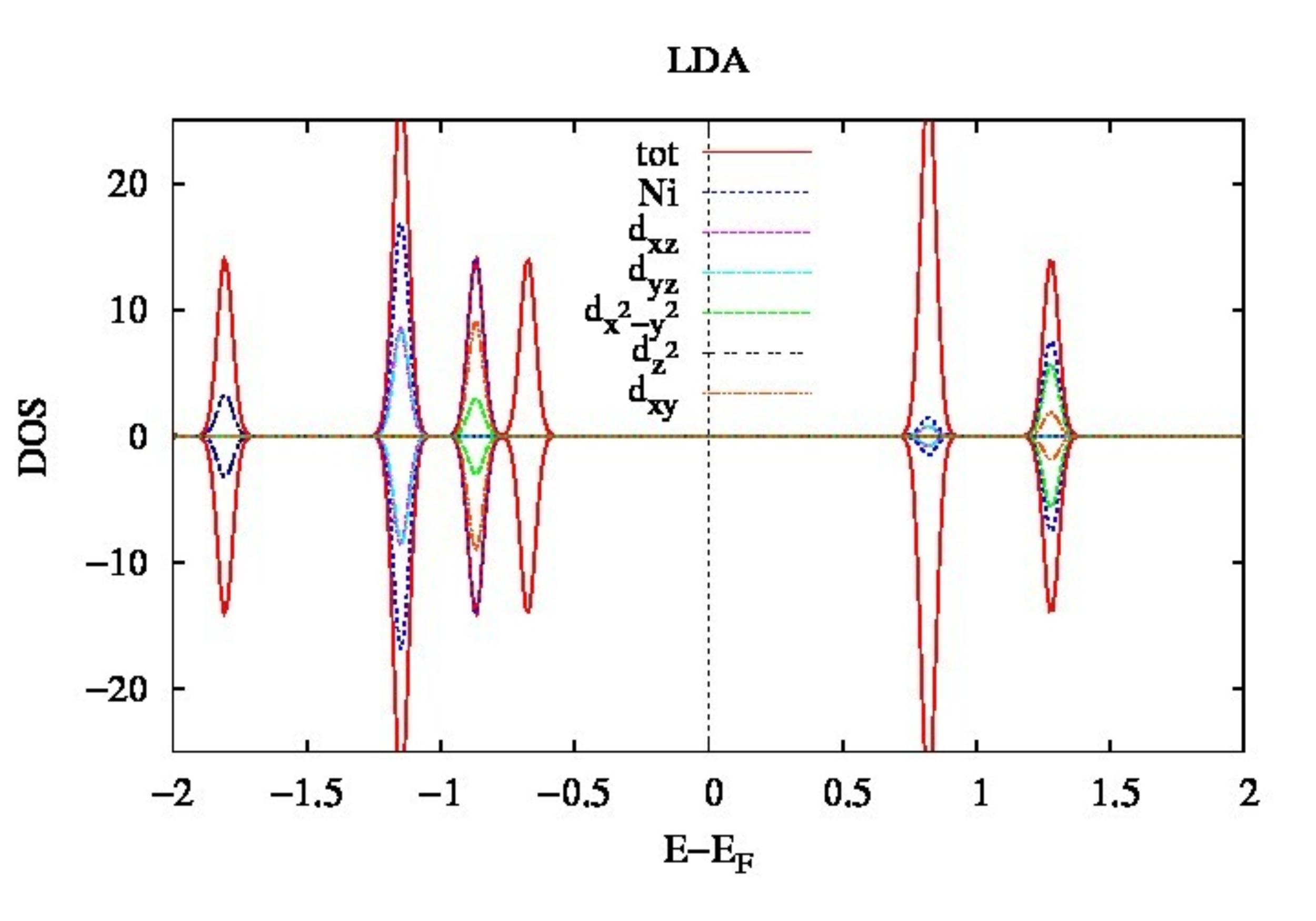}
\includegraphics[width=0.3\textwidth,angle=270]{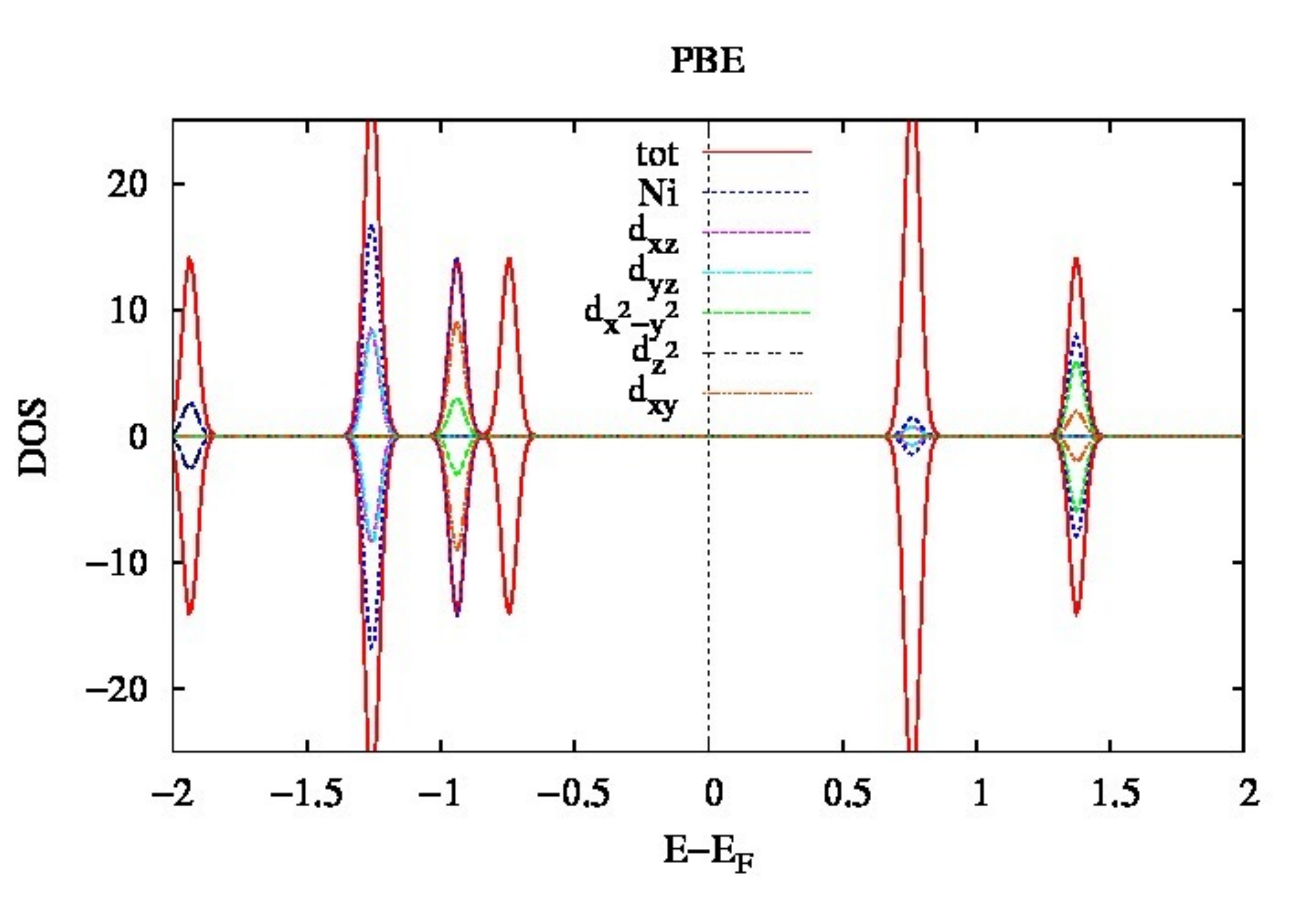}
\end{center}
\begin{center}
\caption{(Color online) Density of states of NiPc with LDA (left) and PBE (right).}
\label{fig15}
\end{center}
\end{figure}

From the NiPc DOS functions shown in Fig.\ \ref{fig15}, it can be noted that there is no difference between
majority and minority spin densities. 
The electron pairing that occurs in this close-shell system causes equal positions and the same orbital order 
for both spins.
Metal orbitals forming occupied orbitals are found below the HOMO, while at the same time there are only very small 
Ni $d_{xz,yz}$ contributions to the LUMO orbitals giving a ligand -- ligand character to HOMO -- LUMO with
the same band gap of $1.5$ eV for both spins. This value is smaller than the one obtained with the Hartree-Fock (HF) method,
2.41 eV \cite{Bi2}, but larger than the LDA prediction, $0.7$ eV, for the monolayer and crystal monoclinic phase 
in the same work. Our band gap is very close to the estimation with the VWN potential of $1.47$ eV \cite{Me}, but that
study also reported that the LUMO corresponds to a $b_{1g}$ orbital, like a later PBE evaluation \cite{NM}. 
Regarding the band gap value and LUMO+1's identity, our result agrees with more recent GGA+U \cite{mugarza12} 
and Spectroscopy-Oriented Configuration Interaction (SORCI) calculations \cite{bruder10}.

\begin{figure}[h!]
\begin{center}
\includegraphics[width=0.3\textwidth,angle=270]{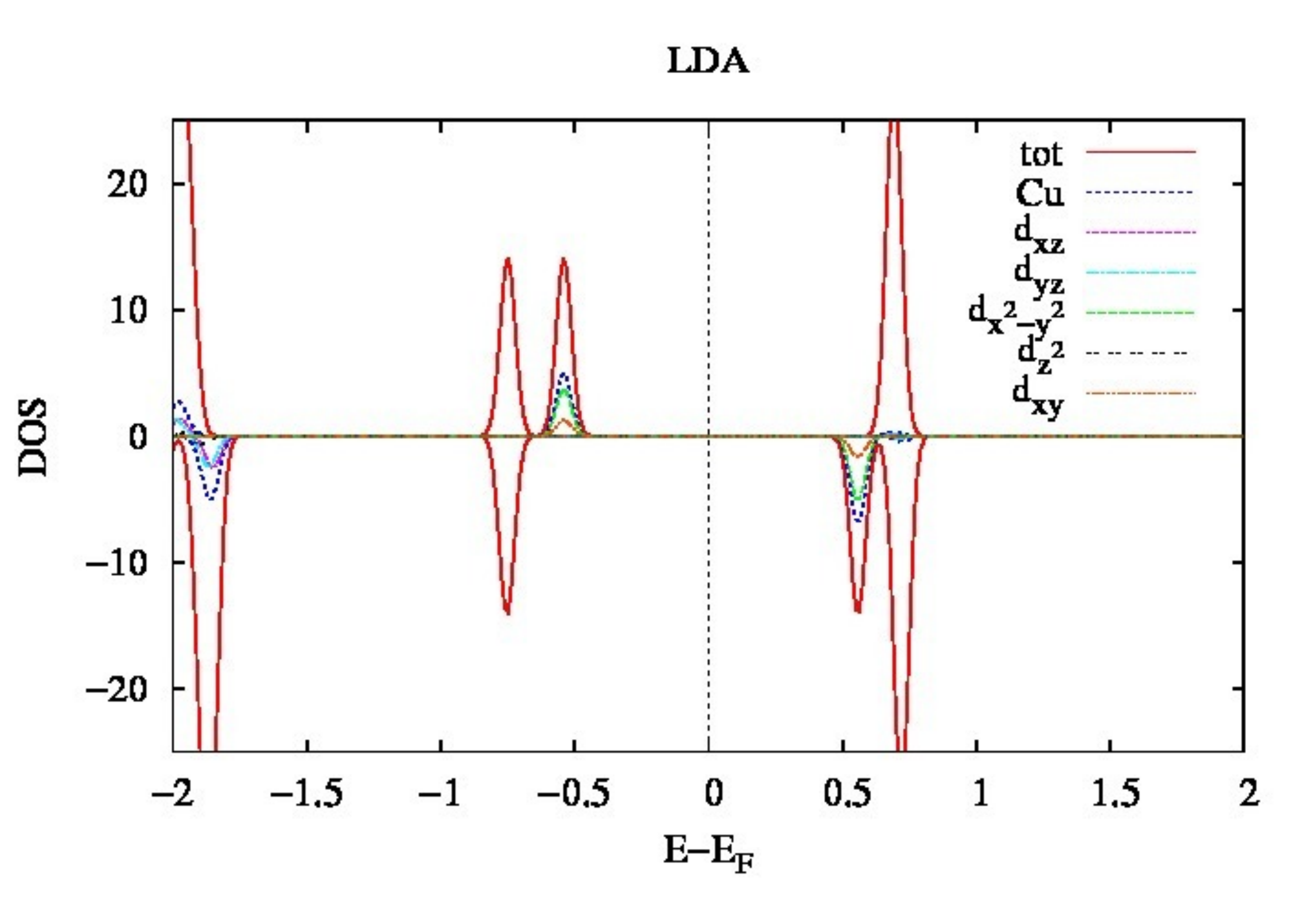}
\includegraphics[width=0.3\textwidth,angle=270]{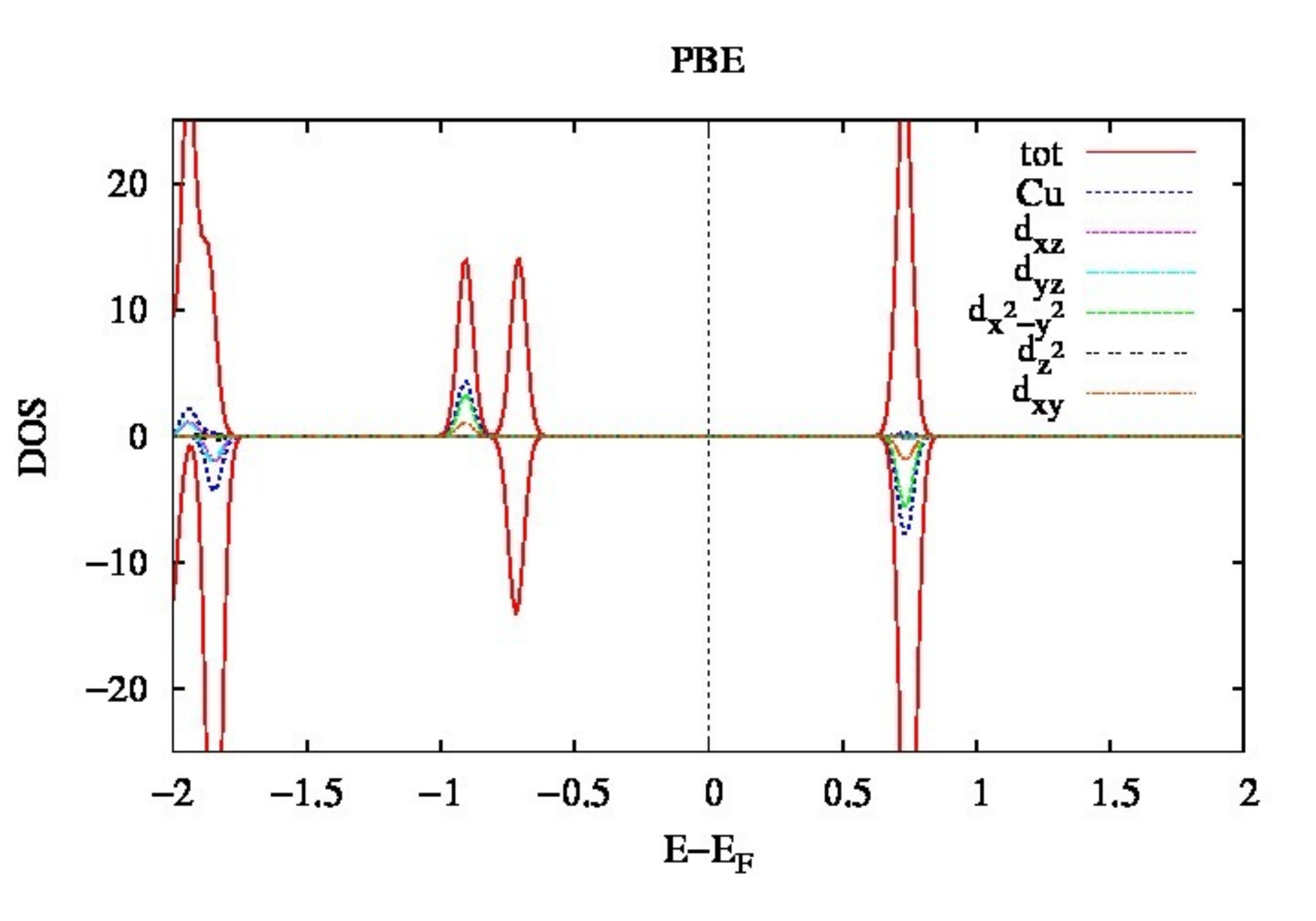}
\end{center}
\begin{center}
\caption{(Color online) Density of states of CuPc with LDA (left) and PBE (right).}
\label{fig16}
\end{center}
\end{figure}

The DOS functions of CuPc (Fig.\ \ref{fig16}) indicates that splitting between $b_{1g}\uparrow$ and
$b_{1g}\downarrow$ increases when the PBE functional is used, making the $a_{1u}$ orbital the HOMO and leaving the 
$b_{1g}\downarrow$ as LUMO. This up- and down-shift of occupied and unoccupied orbitals, respectively, is consistent with 
the expected reduction of the self-interaction error for localized levels. 
The ordering agrees with recent GGA+U calculation including the $1.4$ eV band gap \cite{mugarza12}, and
differs from SORCI results \cite{bruder10}, and previous VWN predictions \cite{Me}. 

\begin{figure}[h!]
\begin{center}
\includegraphics[width=0.3\textwidth,angle=270]{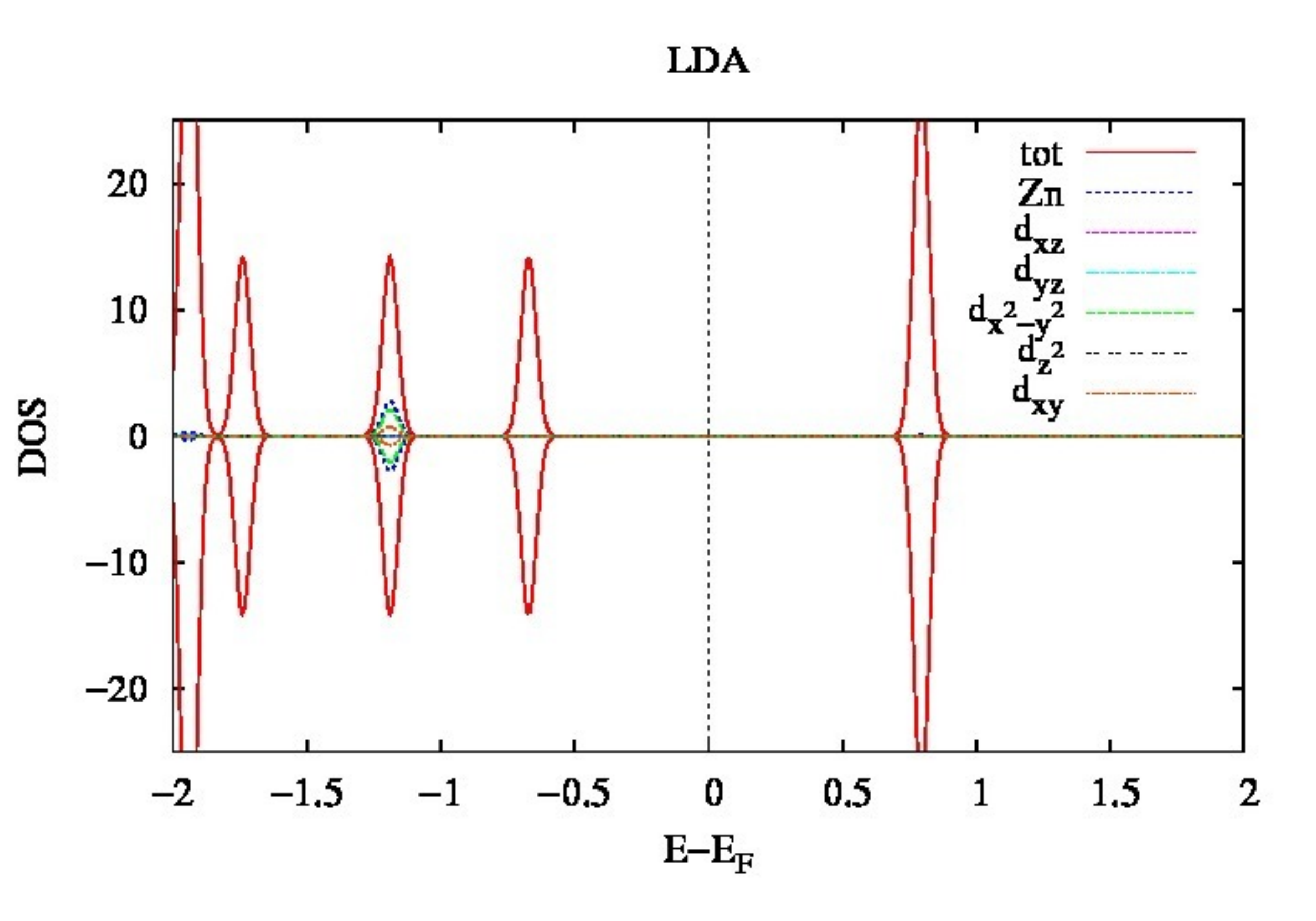}
\includegraphics[width=0.3\textwidth,angle=270]{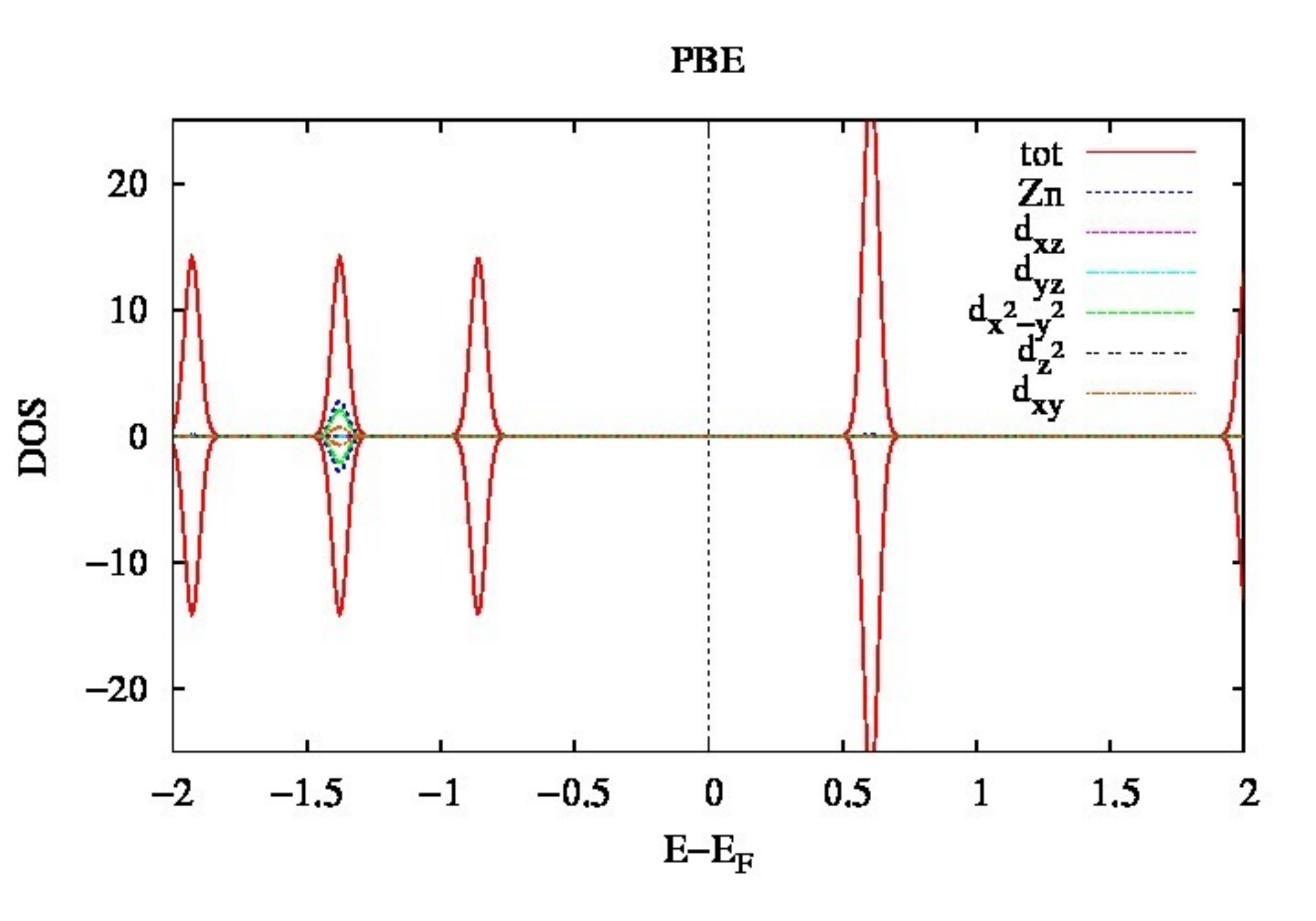}
\end{center}
\begin{center}
\caption{(Color online) Density of states of ZnPc with LDA (left) and PBE (right).}
\label{fig17}
\end{center}
\end{figure}

\begin{figure*}
\includegraphics[width=0.3\textwidth]{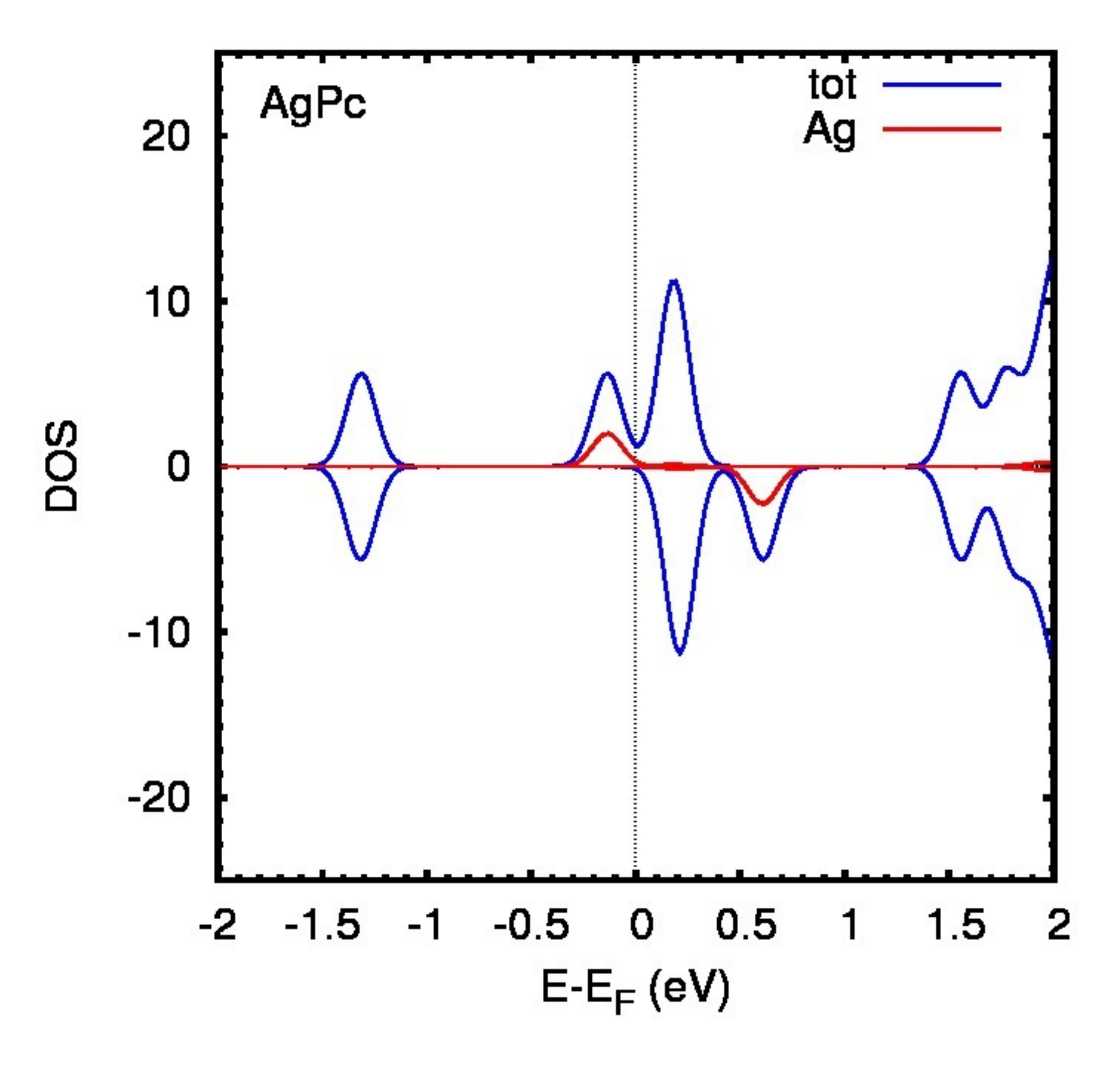}
\caption{(Color online) DOS of AgPc.}
\label{fig18}
\end{figure*}

The last two structures we discuss are ZnPc and AgPc. For ZnPc, the Zn metal states overlap with 
C and N states at $−1.2$ eV for the majority and minority states,
see Fig.\ \ref{fig17}. 
The C and N contributions to the states are observed above $-1.2$ eV, specifically at $-0.75$ eV for C orbitals and at $0.75$ eV for C and N orbitals.
The band gap is $1.5$ eV, similar to the CaPc molecule, but smaller than a previously reported value \cite{KAN}.
For AgPc, the Ag states mix with the N and C states at $-0.1$ eV for the majority states, and at $0.7$ eV for the minority states.
The pure C states prevail at $-1.3$ eV, C and N states at $0.2$ eV and above $1.4$ eV;
thus it is difficult to decide whether AgPc becomes metallic or semiconducting, see Fig.\ \ref{fig18}.

The band gaps for F$_{16}$MPc and for MPc are given in Tab.\ \ref{table2}.
The gap of each F$_{16}$MPc is less than or equal to the gap of the corresponding MPc, which can be correlated
to the fact that fluor is more electronegative, thus the electronic density in the central region of the molecule
is more strongly attracted to the surrounding shell.

\section{Vibrational properties of metal phthalocyanines}

The comparison of the calculated frequencies with experimental data of ZnPc indicates a semi-quantitative 
agreement for the main characteristics of Raman and infrared (IR) spectra (Tab.\ \ref{vibfreq1}).
The experimental Raman spectra presented in Fig.\ 7 from the work of Tackley et al.\ \cite{Tackley} displays 
three bands which are more intense than the rest; our results coincide with their positions within 10 cm$^{-1}$
with the exception of the most intense band at 1505 cm$^{-1}$. 

\begin{table}[h!]
\begin{center}
\caption{Frequencies (cm$^{-1}$) of relevant vibrational modes of ZnPc, and comparison with CuPc} 
{
\renewcommand{\arraystretch}{1.2}
\begin{tabular}{|c|ccc|c|}
\hline
	& \multicolumn{3}{c}{ZnPc}	&	CuPc 	\\
\hline
	& Experimental$^a$ & Calculated$^a$	& Calculated$^b$	& Calculated$^c$	\\
 Raman	& 		&	  	&		&		\\
	&	1505	&	1517 	&	1566	&	1609	\\
	&	1447	&	1438 	&	1448$^d$&	1491	\\
	&	1340	&	1362 	&	1347$^d$&	1395	\\
	&	1144	&	1004 	&	-	&	1351	\\
	&		& 	1134 	&	-	&	1169	\\
	&		&	1278	&	-	&	1035	\\
	&	 747	&	 735 	&	737	&	763	\\
 IR	& 		&	  	&		&		\\
	&	 727	&	 703	&	706 	&	730	\\
	&	 752	&	 741	&	- 	&	769	\\
	&	 780	&	 760	&	- 	&	786	\\
\hline
\end{tabular}
}%
\label{vibfreq1}
\end{center}
\hspace{0.0cm}{\small $^a$From reference \cite{Tackley},}
\hspace{0.0cm}{\small $^b$Our CA/DZP calculation}\\
\hspace{0.0cm}{\small $^c$Our B3LYP/6-31G** calculation}\\
\hspace{0.0cm}{\small $^d$Sorted only by numerical value. See text for explanation.}
\end{table}

Our calculated frequencies of 1448 cm$^{-1}$ and 1347 cm$^{-1}$ 
are close to the experimental bands at 1447 cm$^{-1}$ and 1340 cm$^{-1}$, however,
their displacement eigenvectors have more similarities with those found in
previously reported vibrational modes at 1438 cm$^{-1}$ and 1278 cm$^{-1}$,
respectively (see supplementary information of \cite{Tackley}). The IR and Raman nature of each
vibrational mode considered here is compared to and confirmed through the displacement vector analysis of 
the calculated results for CuPc, also shown in Tab.\ \ref{vibfreq1}.
The match between each vibrational mode type (IR) between ZnPc and CuPc give us the confidence to extrapolate these
results to the rest of the MPc studied here.  

\begin{table}[ht!]
\begin{center}
\caption{Frequencies (cm$^{-1}$) of relevant vibrational modes for MPc}
{
\renewcommand{\arraystretch}{1.2}
\begin{tabular}{|c|cccccccccc|}
\hline
 Mode description       & Sc 	 &	Ti    &	V   &	Cr &	Mn  &	Fe &	Co &	Ni &	Cu &	Zn    \\
\hline
Metal out-of-plane  	&    85    & 	122   &	108   &	144   &	54    &	110   &	168   &	184   &	148   &	110   \\
Asymm. breathing 	&    --    &	193   &	205   &	201   &	171   &	175   & 191   &	197   &	174   & 159   \\
Symm. breathing  	&    --    &	--    &	262   &	262   &	229   &	--    & 260   &	255   &	259   &	264   \\
Ip$^a$ N--M--N bending 	&    194   &	203   &	209   &	210   &	202   &	208   & 243   &	243   &	220   &	218   \\
Op$^b$ asymm. M--N$_4$ bending&  --   &	279   &	--    &	275   &	--    &	243   & 314   &	--    & 250   &	223   \\
Op$^b$ symm. M--N$_4$ bending & 342    & 351   &	381   &	345   &	--    & 361   & 372   &	394   &	331   &	226   \\
\hline
\end{tabular}
}%
\label{vibfreq2}
\end{center}
\hspace{0.0cm}{\small $^a$ Ip = in-plane},
\hspace{0.0cm}{\small $^b$ Op = out-of-plane}
\end{table}

\begin{figure}[h!]
\begin{center}
\leavevmode
\includegraphics[width=0.3\textwidth]{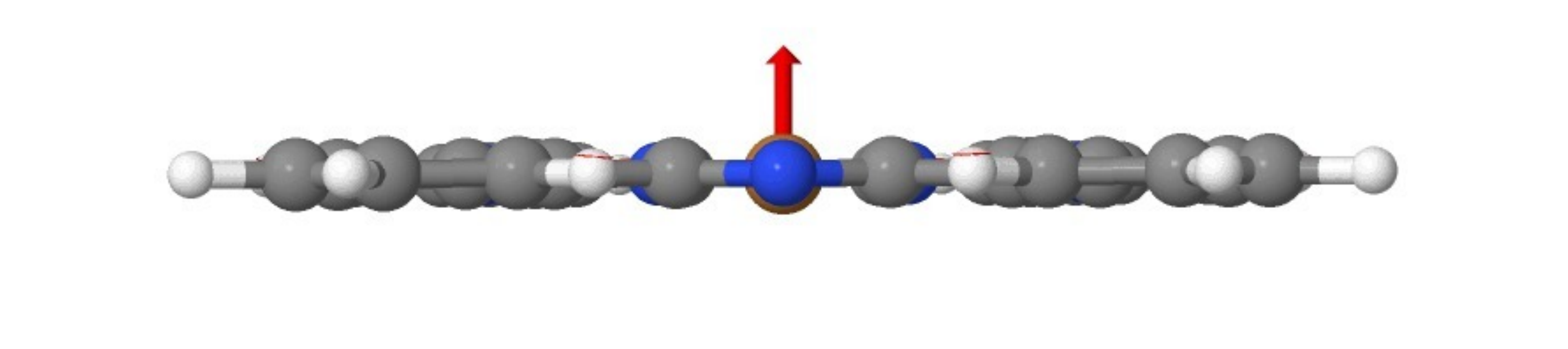}
\includegraphics[width=0.3\textwidth]{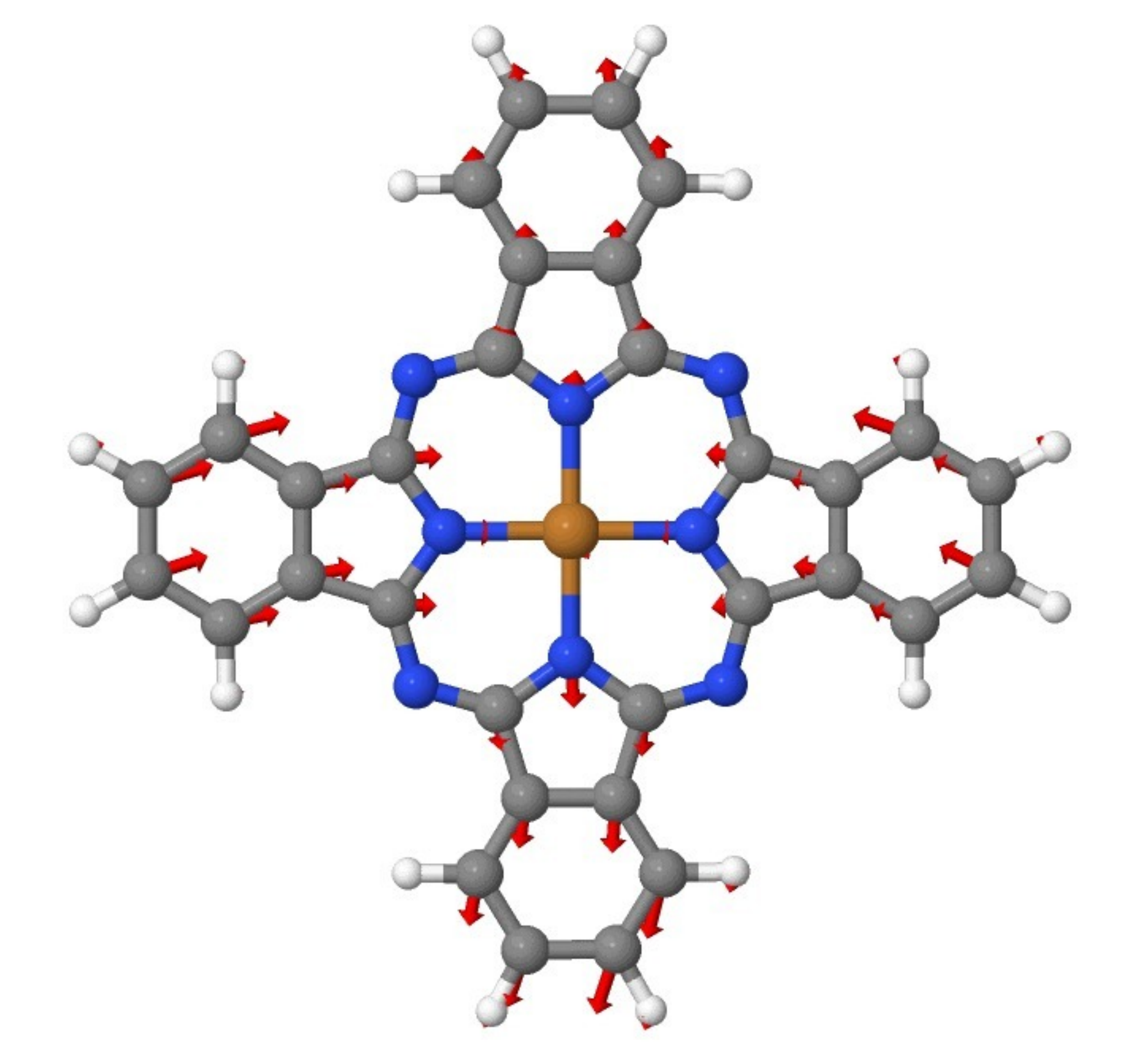}
\hspace{0.0cm}{(a)}\hspace{6.0cm}{(b)}\\
\vspace{0.25cm}
\includegraphics[width=0.3\textwidth]{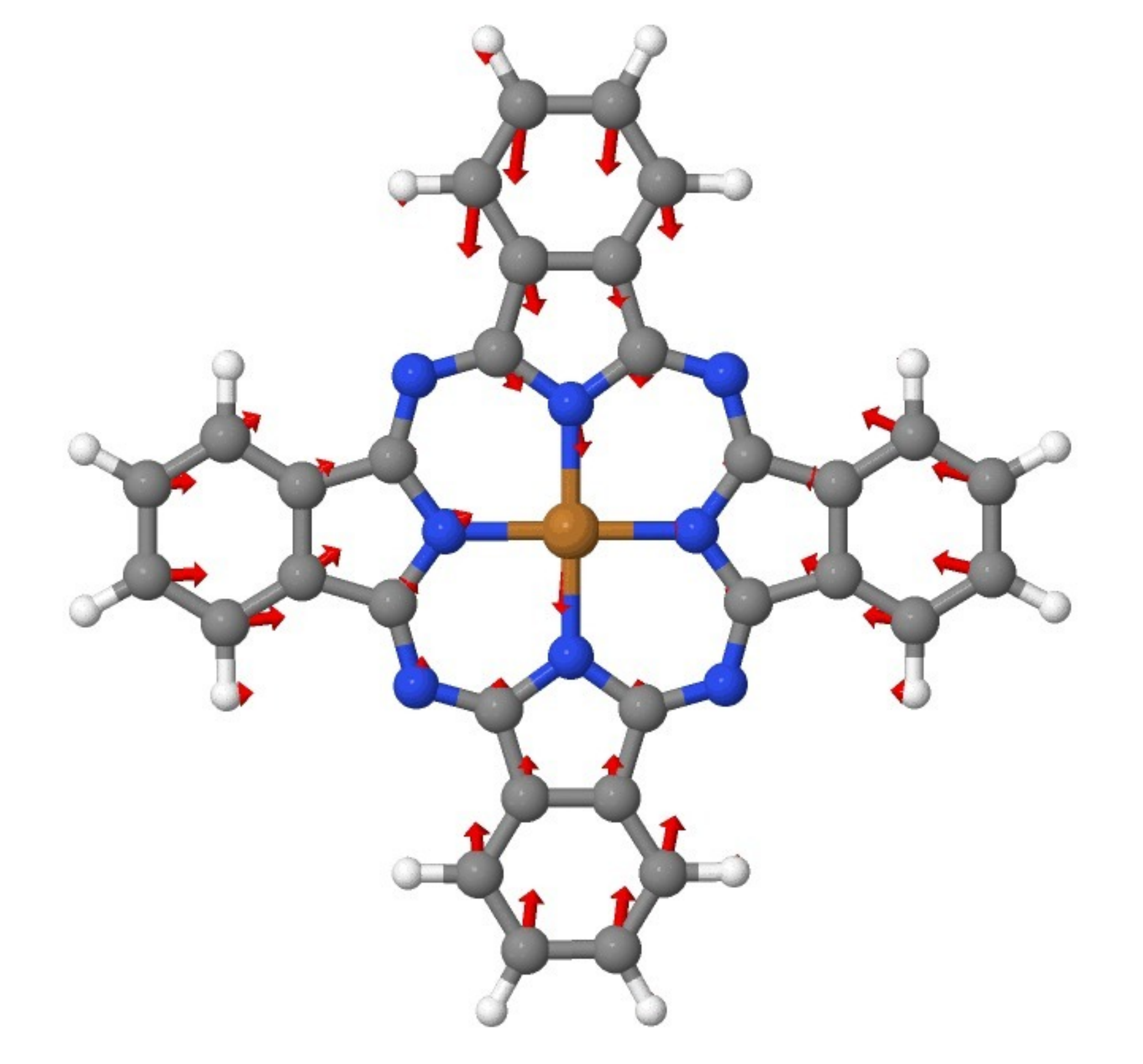}
\includegraphics[width=0.3\textwidth]{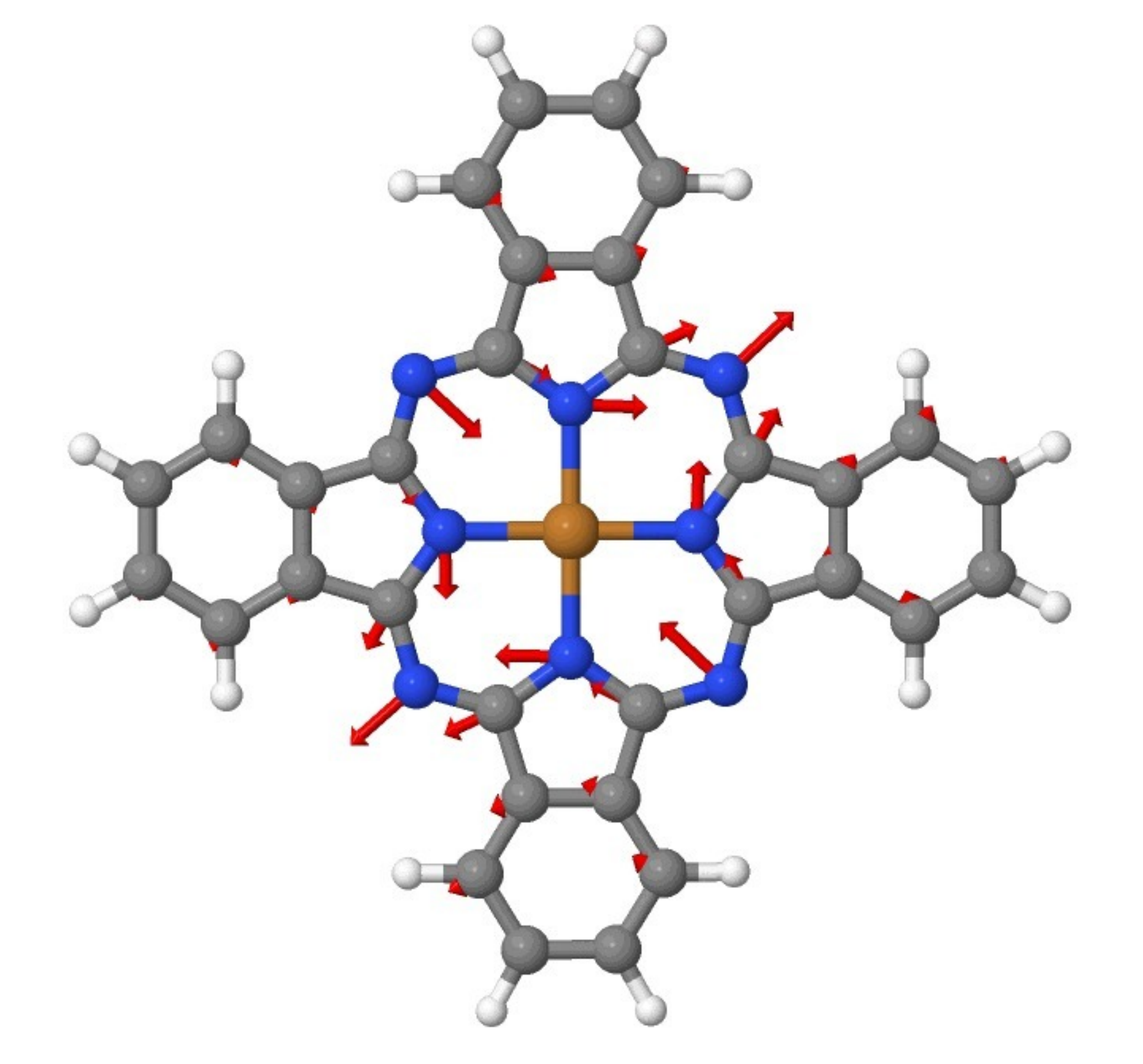}
\hspace{0.0cm}{(c)}\hspace{4.5cm}{(d)}\\
\vspace{0.25cm}
\includegraphics[width=0.3\textwidth]{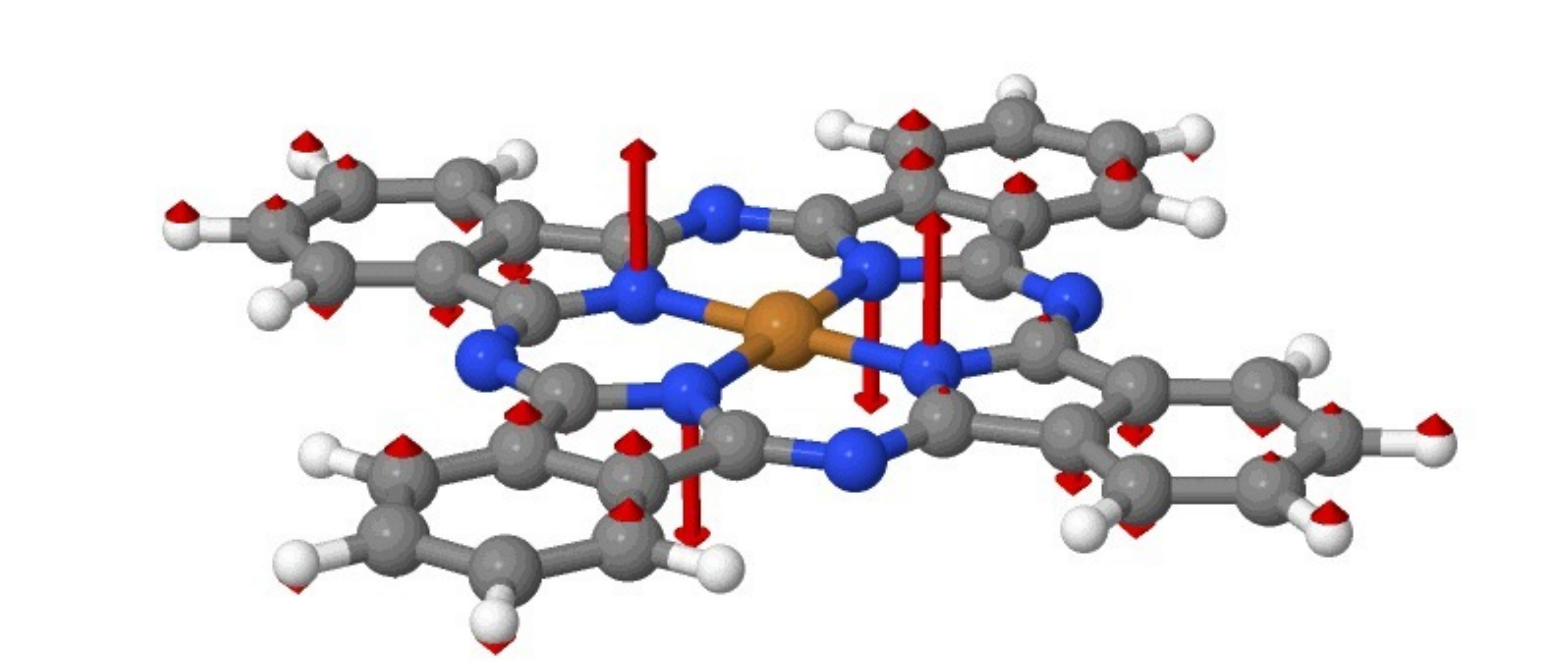}
\includegraphics[width=0.3\textwidth]{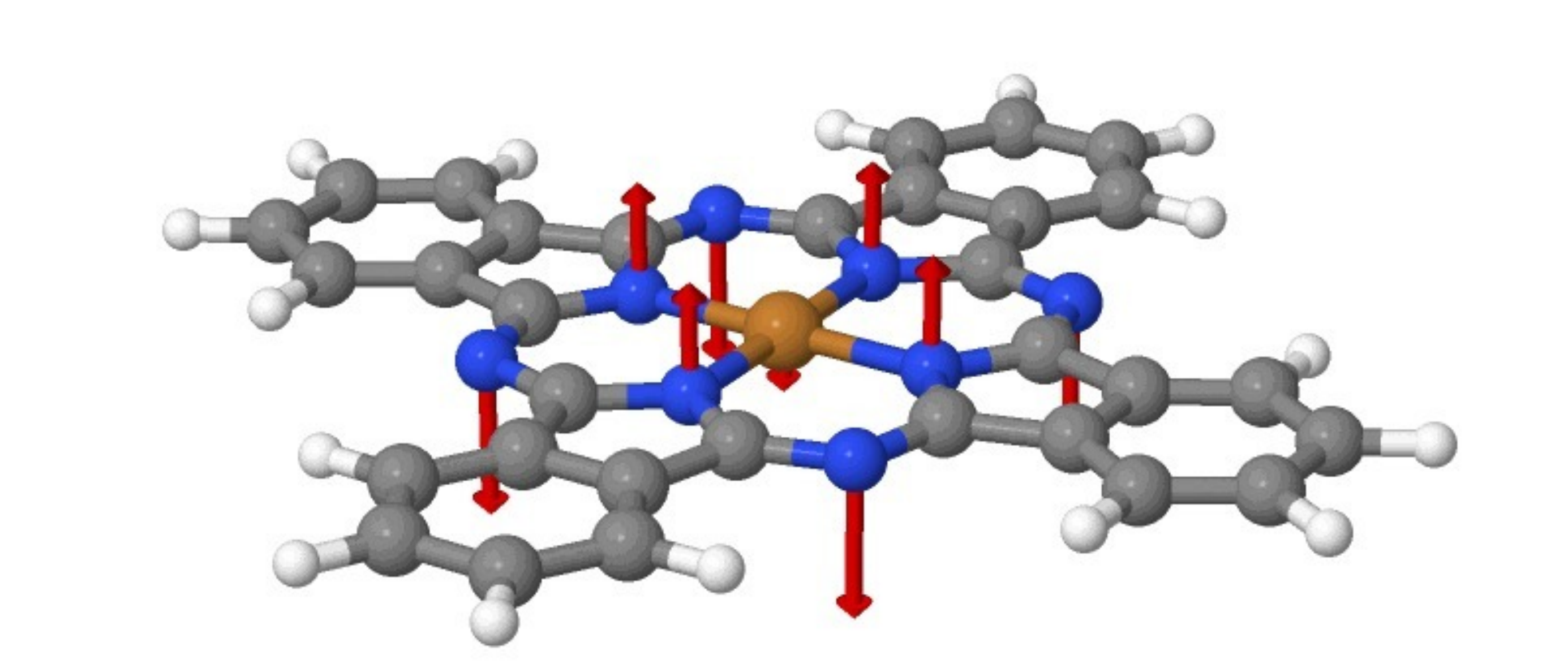}
\hspace{0.0cm}{(e)}\hspace{6.0cm}{(f)}\\
\end{center}
\begin{center}
\caption{(Color online) Selected vibrational modes of CuPc also observed in other MPc, and described as metal out-of-plane
displacement (a), asymmetric (b) and symmetric (c) breathing, in-plane N--M--N bending (d), out-of-plane asymmetric (e) and symmetric (f) M--N$_4$ bending.}
\label{vectors1}
\end{center}
\end{figure}

\begin{center}
\begin{table}[ht!]
\caption{Frequencies (cm$^{-1}$) of relevant vibrational modes for MPc}
{
\renewcommand{\arraystretch}{1.2}
\begin{tabular}{|c|cccccccccc|}
\hline
 Mode description       & Sc 	 &	Ti 	 &	V   	 &	Cr 	 &	Mn 	 &	Fe	&	Co 	 &	Ni 	 &	Cu 	 &	Zn       \\
\hline
M--N$_2$ stretching-bending   & 326  & 341   	& 352	 & 	357   &	312   &	309	& 389	& 345	& 301	& 241	\\
M--N$_2$ stretching-bending   & 329  & 342   & 358     &	363   &	316   &	363     & 401   & 351   & 302   & 245   \\
M--N$_4$ asymm. stretching& 544  & 545  &	542  &	546  &  537  &	511  &	546  &	542  &	539  &	533  \\
M--N$_4$ symm. stretching	& 576  & 574  &	572  &	576  &	569  &	575  &	581  &	578  &	577  &	576  \\
M--N$_4$ asymm. stretching& 1325 & 1183 &	1205 &	1357 &	1303 &	1342 &	1359 &	1362 &	1351 &	1347 \\
M--N$_4$ symm. stretching	& 1366 & 1337 &	1387 &	1401 &	1348 &	1375 &	1407 &	1407 &	1403 &	1409 \\
\hline
\end{tabular}
}%
\label{vibfreq3}
\end{table}
\end{center}

\begin{figure}[h!]
\begin{center}
\leavevmode
\includegraphics[width=0.3\textwidth]{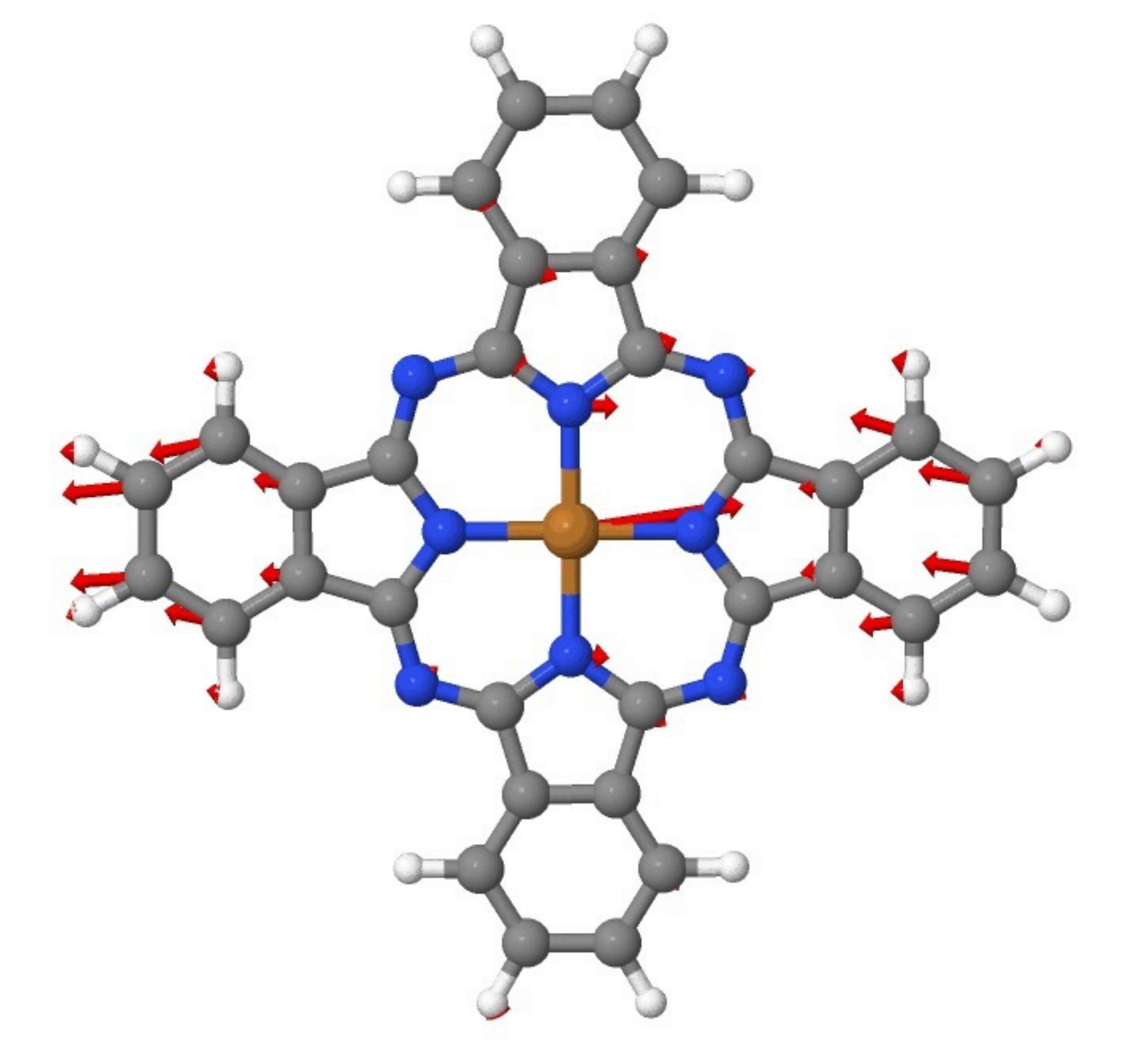}
\includegraphics[width=0.3\textwidth]{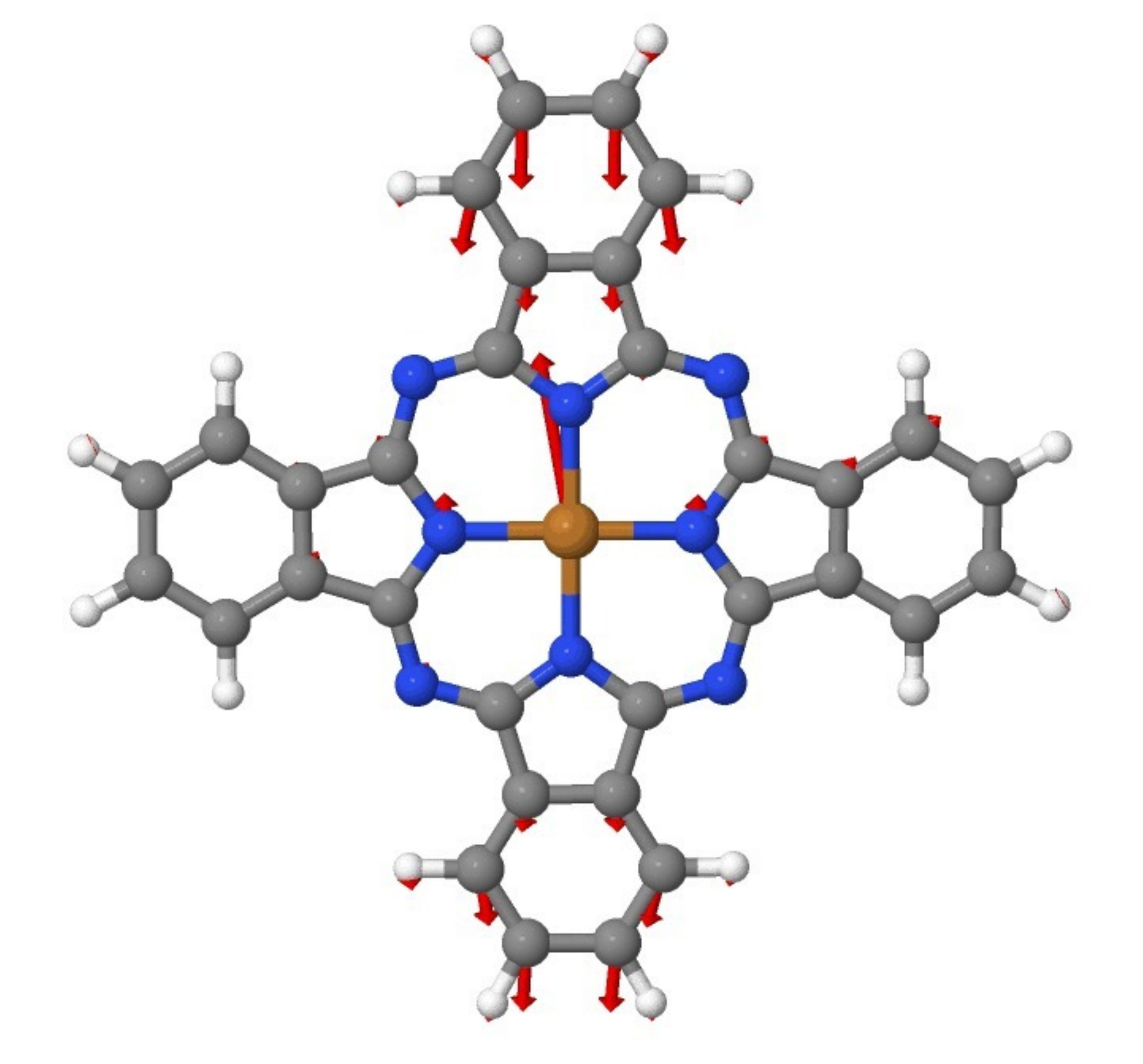}
\hspace{0.0cm}{(a)}\hspace{5.0cm}{(b)}\\
\vspace{0.25cm}
\includegraphics[width=0.3\textwidth]{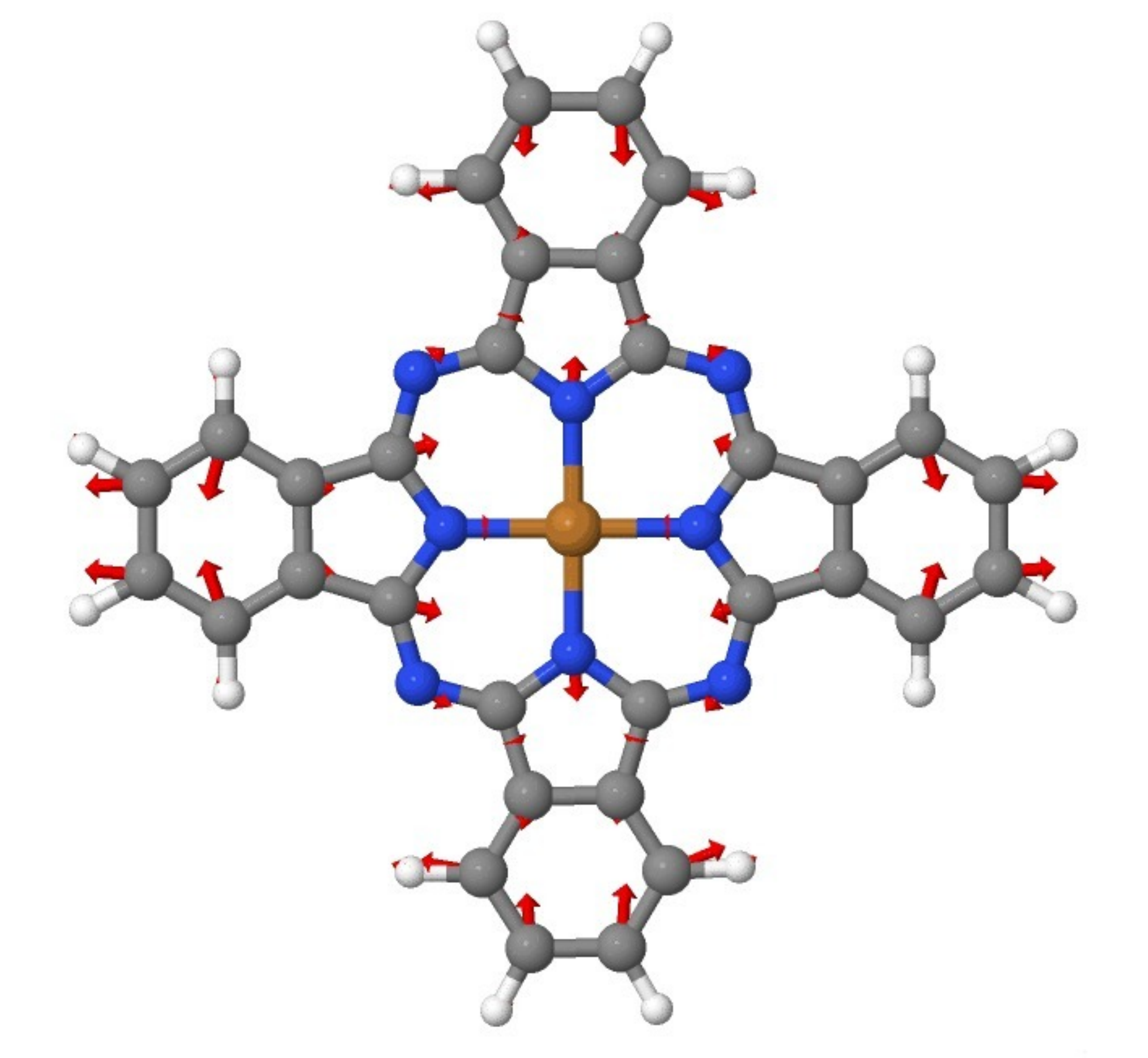}
\includegraphics[width=0.3\textwidth]{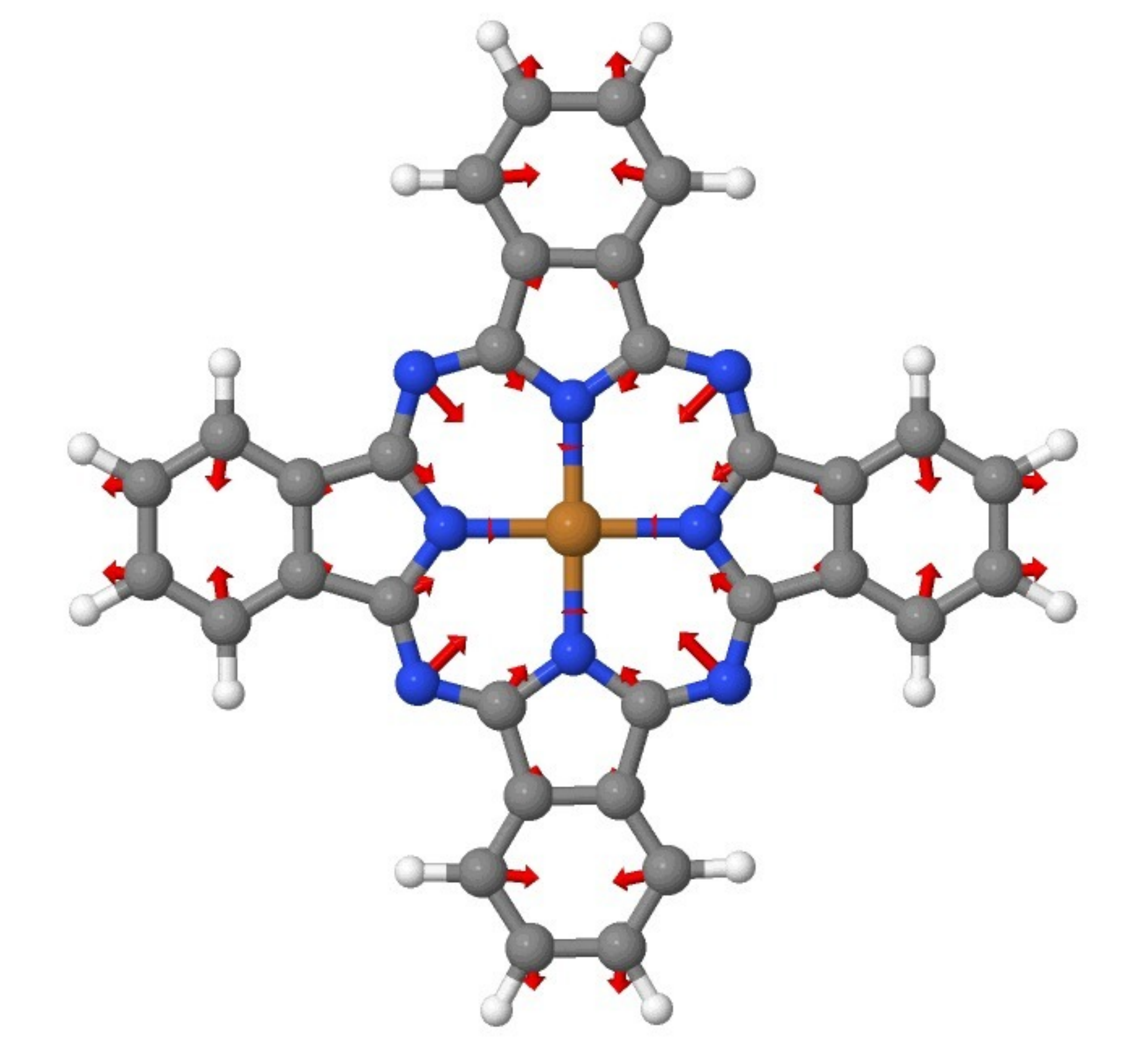}
\hspace{0.0cm}{(c)}\hspace{5.0cm}{(d)}\\
\vspace{0.25cm}
\includegraphics[width=0.3\textwidth]{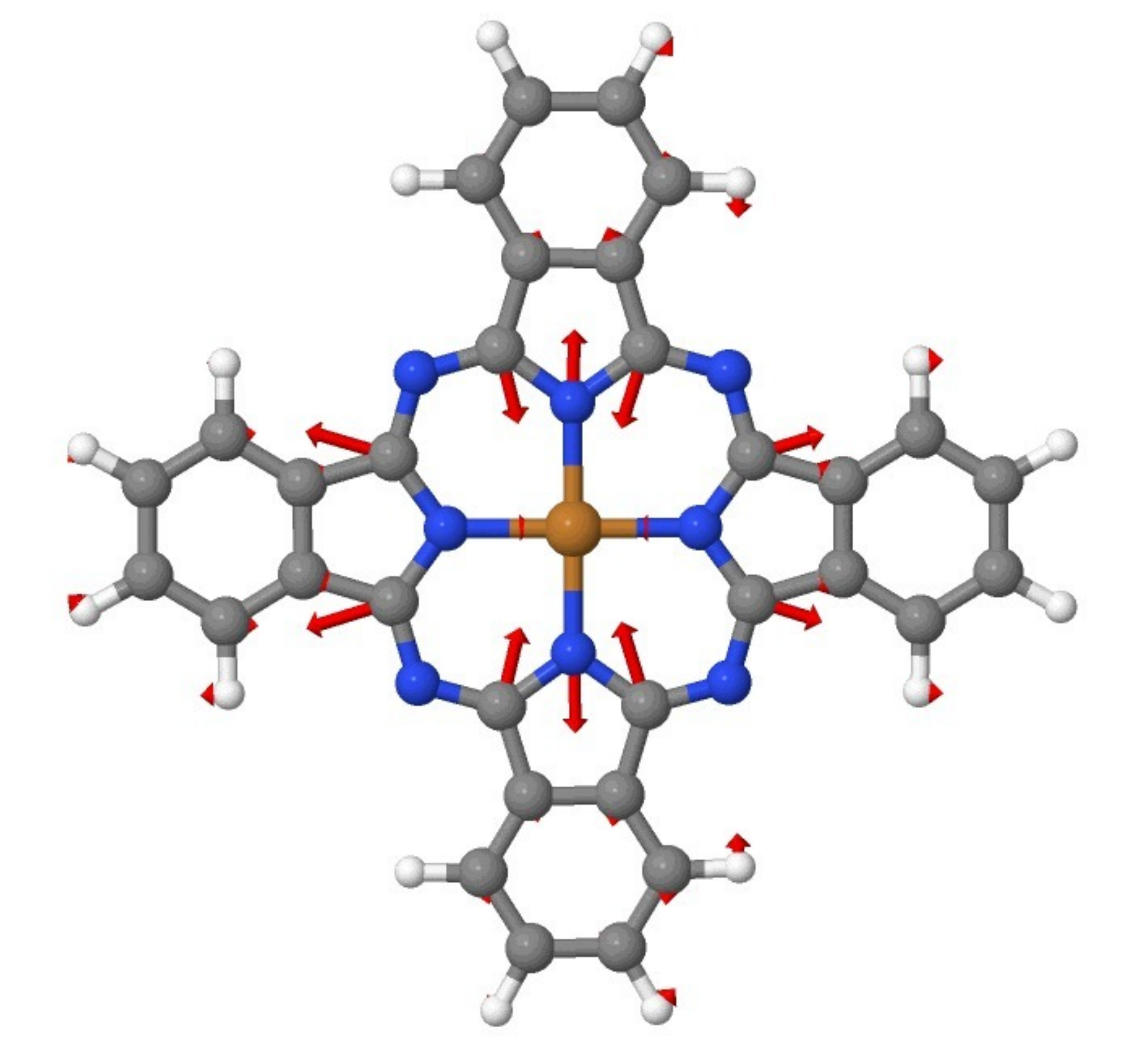}
\includegraphics[width=0.3\textwidth]{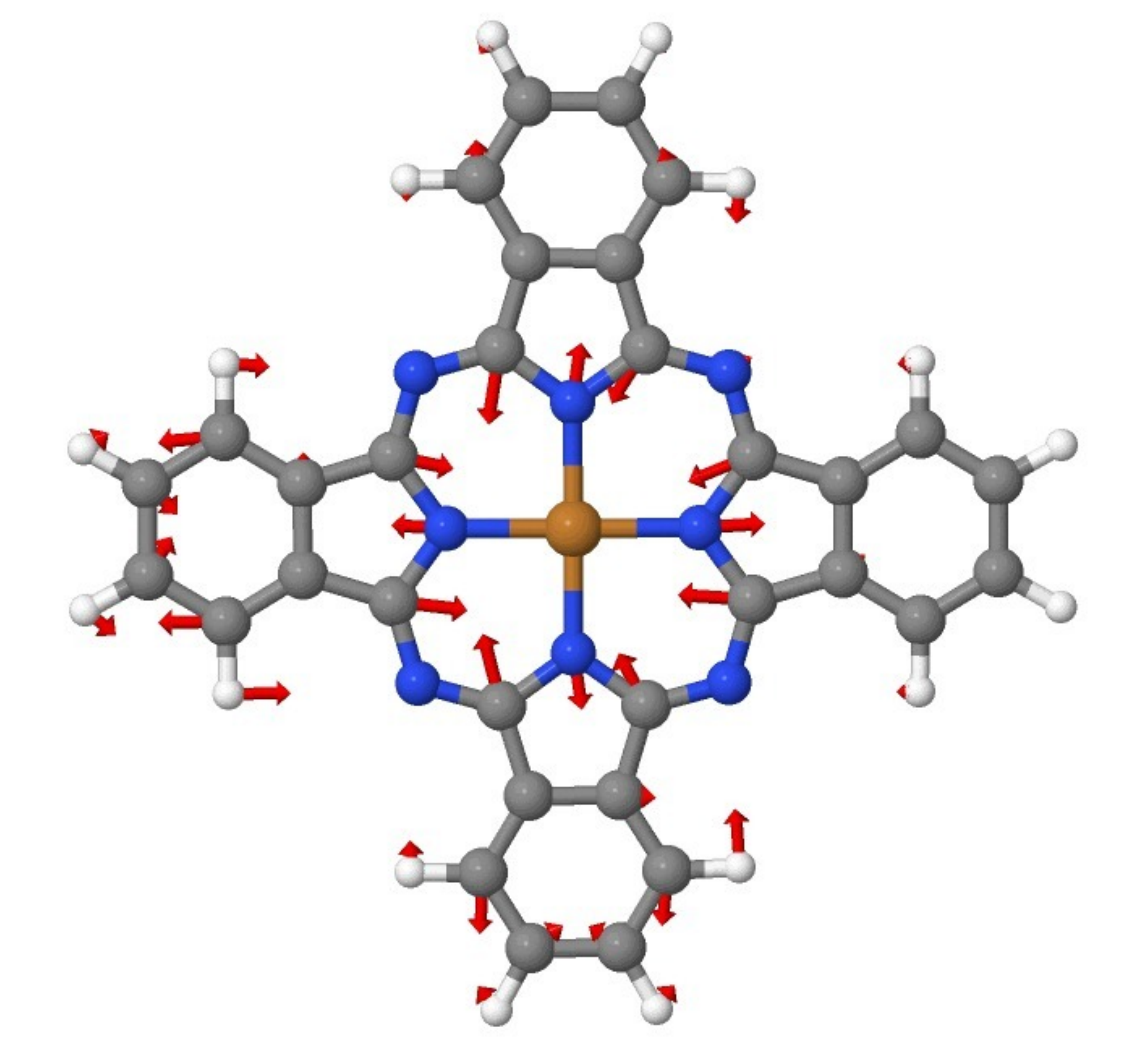}
\hspace{0.0cm}{(e)}\hspace{6.0cm}{(f)}\\
\end{center}
\begin{center}
\caption{(Color online) Selected vibrational modes of CuPc, also observed in other MPc. M--N$_2$ stretching-bending, (a) and (b); M--N$_4$
asymmetric ((c), (e)) and symmetric ((d), (f)) stretching.}
\label{vectors2}
\end{center}
\end{figure}

\begin{figure}[h!]
\begin{center}
\leavevmode
\includegraphics[width=0.3\textwidth,angle=270]{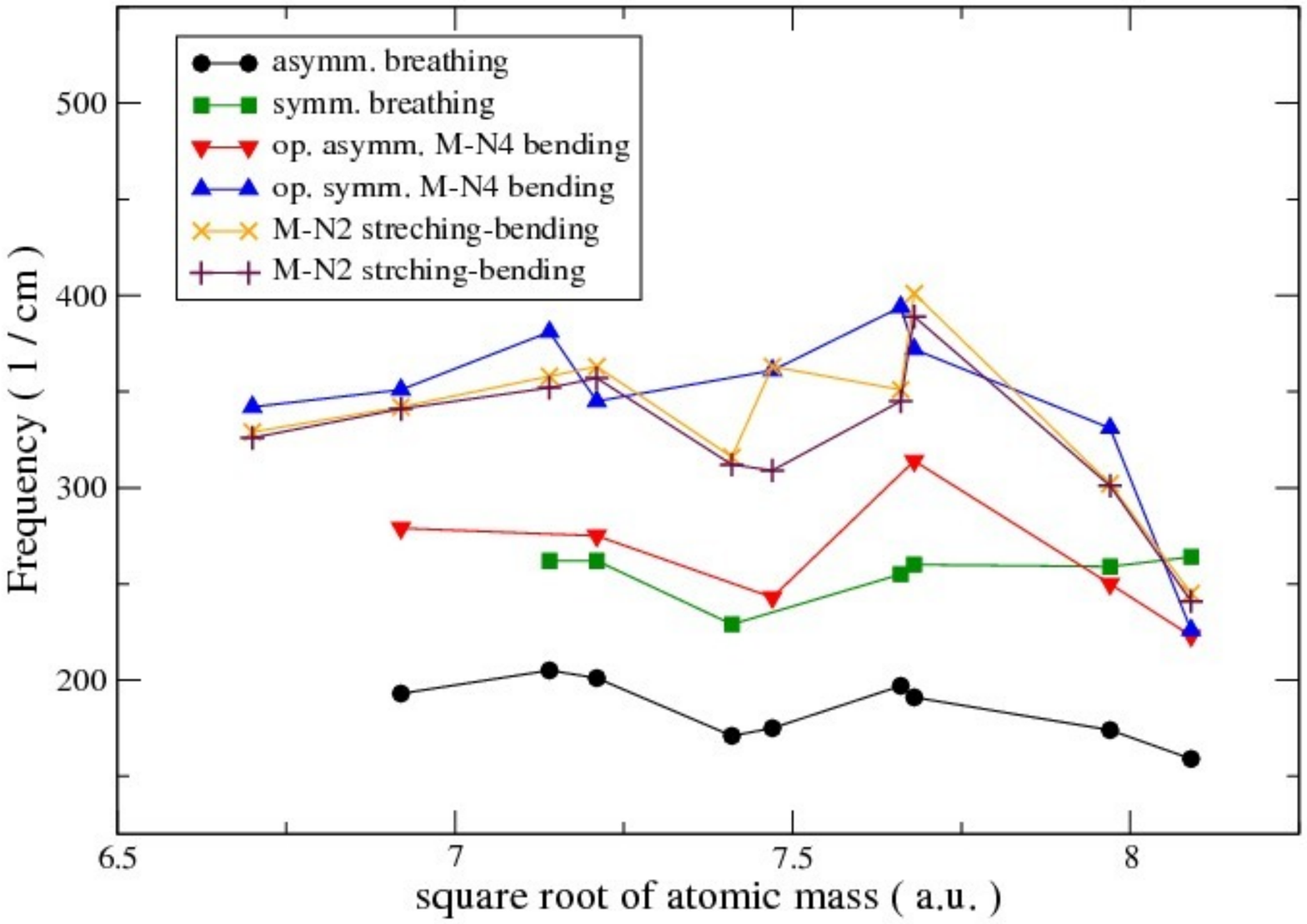}
\includegraphics[width=0.3\textwidth,angle=270]{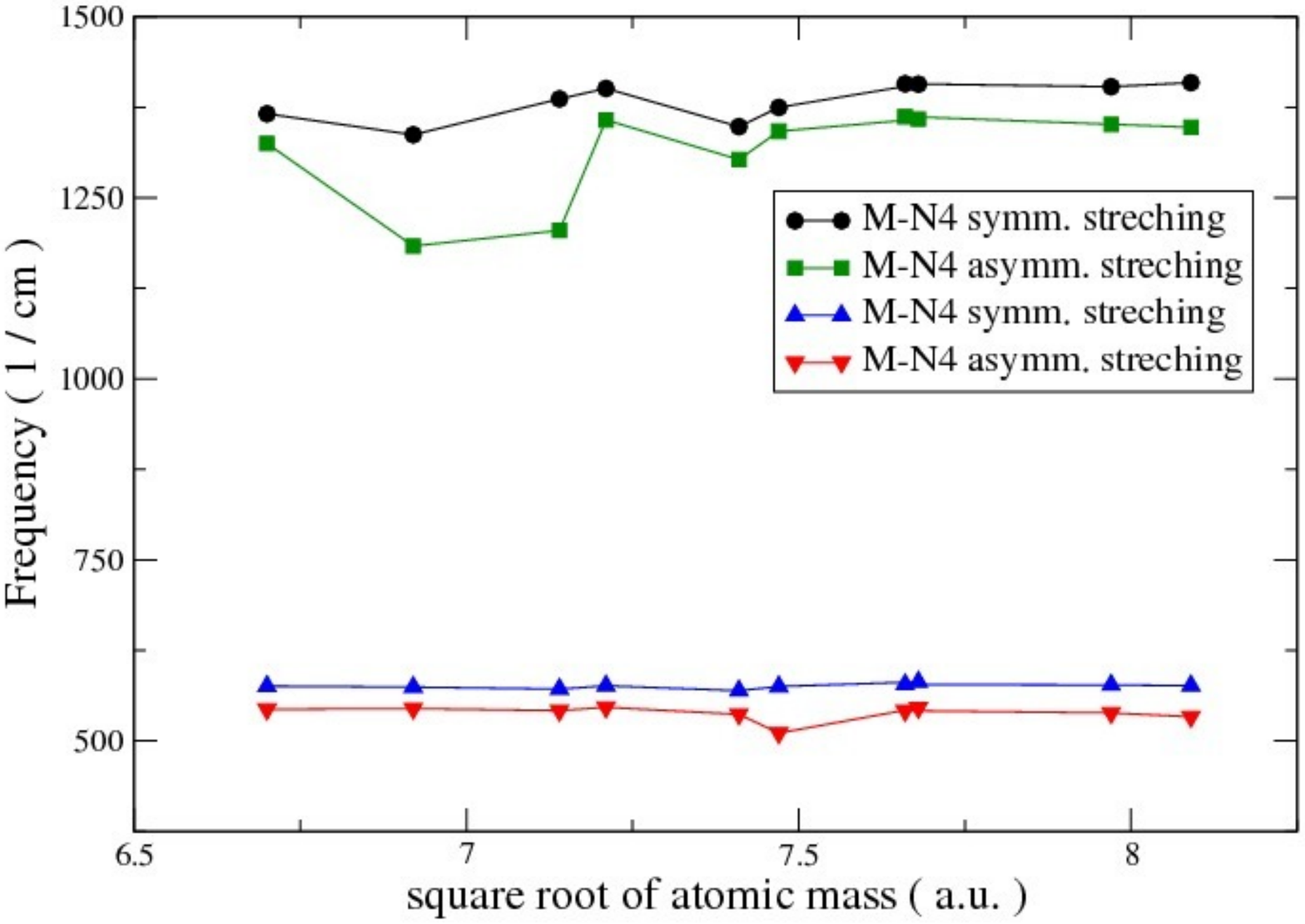}
\end{center}
\begin{center}
\caption{(Color online) Frequencies of selected vibrational modes of MPc}
\label{compfmode2}
\end{center}
\end{figure}

Tables \ref{vibfreq2} and \ref{vibfreq3} show calculated frequencies of the MPc complexes for relevant 
vibrational modes such as breathing, and for those which involve atomic displacement near the metal.
Displacement vectors for such modes are schematically presented in Figs.\ \ref{vectors1} and \ref{vectors2}.
As expected, symmetric vibrational modes are found at higher frequencies than asymmetric modes, but their frequency 
shift, when the metal center is varied, depends on the particular mode. Symmetric and asymmetric breathing movements of the entire 
molecule show similar magnitudes of frequency shifts, the main difference being that frequencies of asymmetric breathing modes
increase when going from Zn to Co, while the opposite tendency is found for the symmetric modes (see Fig.\ \ref{compfmode2}).
This behavior may be explained by the release of stress around the metal inside the ring when it opens simultaneously
in all directions.

In the case of the M--N$_4$ stretching modes, it is apparent from Fig.\ \ref{compfmode2} that there is
a consistent increase from Zn to Co or Ni for the asymmetric type. The high-frequency ($>$ 1180 cm$^{-1}$) modes 
of M--N$_4$ stretching have larger shifts. Their frequencies decrease from Fe to Co, then they increases again for Cr.
These variations are not present for low frequency ($<$ 600 cm$^{-1}$) M--N$_4$ stretching modes: this
can be understood on the basis of their displacement vectors (Fig.\ \ref{vectors2}), which indicate significant 
deformations of benzene rings and smaller displacements on atoms close to the metal. 
Vibrational modes which involve atomic displacements near the metal show larger shifts. For bendings the magnitude
of the shifts can be tens of cm$^{-1}$. There is a frequency increase from Zn to Co, then a decrease for Fe, and 
an increase again for Cr. This is found for out-of-plane M--N$_4$ bending, and for M--N$_2$ stretching-bending modes. 

In Tab.\ \ref{vibfreq4} we compare our frequency values and the nature of vibrational modes of CuPc with those reported 
in \cite{Wu} for the fluorinated system FCuPc,
particularly for those which were found to have a larger contribution to the energy transfer process.
Frequency values for CuPc become similar to results for FCuPc when the vibrational modes involve
a small number of atoms, and when the atomic displacements are mainly from atoms closer to the metal.
The above trends are also obtained with PBE.

\begin{table}
\begin{center}
\caption{Comparison of frequencies (cm$^{-1}$) of relevant vibrational modes of CuPc and FCuPc} 
{
\renewcommand{\arraystretch}{1.2}
\begin{tabular}{|c|cccc|}
\hline
	FCuPc$^a$	&	175	&	983	&	1445	&	1570	\\
	CuPc$^b$	&	201	&	1158	&	1401 	&	1561    \\
\hline
\end{tabular}
}%
\label{vibfreq4}
\end{center}
\hspace{0.0cm}{\small $^a$From reference \cite{Wu}}\\
\hspace{0.0cm}{\small $^b$Our CA/DZP calculation}\\
\end{table}

\newpage
\pagebreak
\newpage

\section{Summary}
The magnetic and electronic properties of metal phthalocyanines (MPc) and fluorinated metal phthalocyanines (F$_{16}$MPc) 
have been studied by spin density functional theory for the first row-transition metals. CaPc and AgPc were also considered 
for comparison.
A planar geometry with the metal lying inside the inner ring was obtained for all MPc except for the cases M = Ca and Sc, which
are displaced from the plane by 1.12 \AA\ and 0.24 \AA, respectively, in direct relation with their atomic sizes.
The analysis of electronic states shows that the total magnetic moment is caused mainly by the electron 
density localized on the metal and its nearby atoms. 
Furthermore, when the hydrogen atoms are replaced by fluor atoms the magnetic moments do not change. 
The same behavior is found for the band gap values. The M-N bond lengths of F$_{16}$MPc are slightly larger than
the corresponding ones of MPc. 

On the other hand, the analysis of the density of states (DOS) indicates a substantial change in the number, 
position, width and height of the DOS peaks in F$_{16}$MPc in comparison to MPc. The fact that the total
magnetic moment values do not change can be explained by the charge density redistribution between the metal and 
its nitrogen and carbon neighbors.
This rearrangement suggests a high flexibility of
the electronic distribution which may account for the apparent independence of the MPc properties on the metal, also
pointed out in previous works \cite{takacs}. However, the DOS profiles obtained along the studied $d$-row series indicate
different mixing between metal and ligand orbitals, and corresponding energy shifts.

The crystal field approximation cannot account for the orbital mixing occurring in these compounds.
The ligand orbitals $a_{1u}$ and $e_{g}$ have the propensity to appear near the Fermi energy and
they can become HOMO or LUMO when the electronic repulsion shifts localized states on the metal 
or when electron pairing on the metal compensates repulsion between levels for both spins as in NiPc and ZnPc.
Near the middle of the studied $d$ series, repulsion among unpaired electron densities causes different
DFT functionals to disagree with respect to the electronic structure.
However, we have found that in general, LDA results are valuable for the qualitative assessment of the metallic or
ligand character of the HOMO and the LUMO. Comparison with PBE results is useful to determine which
localized metal orbitals are more strongly affected by the self-interaction error.
The observed tendencies may be extrapolated to understand and predict the device performance of particular 
MPc employed in organic solar cells in which the exciton diffusion is affected by the energetical ordering 
of the orbitals \cite{bruder10}.

The band gap shows a steep drop from CaPc to ScPc, i.e., when the $d$ shell is initated. The largest value found in the
$d^2$-$d^4$ interval is for CrPc, and another important decrease is predicted for MnPc.
The semiconducting character is recovered for the $d^7$-$d^{10}$ series with low spin configuration.
The crucial effect of the central metal in the electronic properties of MPc helps to explain recent theoretical results 
about the energetics of their adsorption on an Au(111) surface \cite{zhang11}, as well as the increasing impact of the 3$d$ states
along the ZnPc\textendash MnPc series as determined by energy-loss and photoemission spectroscopy \cite{grobosch10}.
Ongoing research regarding electron transport on these MPc is being carried out in our group.

Based on the vibrational mode analysis, we find that, in general, frequencies increase when going from Zn to Fe if they correspond
to asymmetric modes. There is an increment from either Mn or Fe to Cr. The series V, Ti, and Sc shows a more irregular behavior.
However, there is a consistent frequency decrease from V to Sc for vibrational modes which do not include breathing or metal 
out-of-plane movement.
The vibrational frequencies change a few tens of cm$^{-1}$ when the metal in the central position is varied. 
This is an order of magnitude larger than the shifts provoked by molecular adsorption \cite{Saini}.

\vspace*{-0.3cm}
\subsection*{Acknowledgement}
\vspace*{-0.3cm}
We thank P.\ Schwab for helpful discussions. Financial
support by the Deutsche Forschungsgemeinschaft (through TRR 80), PPPROALMEX-DAAD-Conacyt binational support, Project SEP-CONACYT 152153,
CNS-IPICyT, Mexico and TACC-Texas supercomputer
center for providing computational resources. O. I. A.-F. thanks Camilo Garcia for valuable discussions and CONACyT for a postdoctoral
fellowship. AHR acknowledges support from the Marie-Curie Intra-European Fellowship program.


\end{document}